\documentclass[review]{elsarticle}
\usepackage{amsmath}
\usepackage{lineno,hyperref}

\usepackage{array}
\usepackage{amsfonts}
\usepackage{tipa}
\usepackage{xcolor}
\usepackage{float}
\usepackage{subfig}
\usepackage{booktabs}
\usepackage{multirow}

\graphicspath{ {images/} }

\usepackage{arabtex}
\usepackage{utf8}
\usepackage[utf8]{inputenc}
\setcode{utf8}
\usepackage[export]{adjustbox}

\def\endabstract{\egroup}
\DeclareMathOperator*{\argmaxA}{argmax}
\modulolinenumbers[5]

\journal{Journal of Computer Speech and Language}

\begin{document}

\begin{frontmatter}

\title{Arabic Speech Recognition by End-to-End, Modular Systems and Human}

\author[address1,address2]{Amir Hussein\corref{author}}
\ead{amir@kanari.ai}


\author[address3]{Shinji Watanabe}
\ead{shinji@ieee.org}

\author[address1]{Ahmed Ali}
\ead{amali@hbku.edu.qa}

\address[address1]{HBKU, Qatar Computing Research Institute, Doha, Qatar}

\address[address2]{Kanari AI, Pasadena, California}
\address[address3]{Carnegie Mellon University}

\begin{abstract}
    Recent advances in automatic speech recognition (ASR) have achieved accuracy levels comparable to human transcribers, which led researchers to debate if the machine has reached human performance. Previous work focused on the English language and modular hidden Markov model-deep neural network (HMM-DNN) systems. In this paper, we perform a comprehensive benchmarking for end-to-end transformer ASR, modular HMM-DNN ASR, and human speech recognition (HSR) on the Arabic language and its dialects. For the HSR, we evaluate linguist performance and lay-native speaker performance on a new dataset collected as a part of this study. For ASR the end-to-end work led to 12.5\%, 27.5\% , 33.8\% WER; a new performance milestone for the MGB2, MGB3, and MGB5 challenges respectively. Our results suggest that human performance in the Arabic language is still considerably better than the machine with an absolute WER gap of 3.5\% on average.
\end{abstract}


\end{frontmatter}








\section{Introduction}

Automatic Speech Recognition has shown fast progress recently, thanks to advancements in Deep Neural Network (DNN) which has brought remarkable improvements in reaching human-level performance \cite{deepspeech2}. Traditional ASR systems employ a modular design, with different modules for acoustic modeling, pronunciation lexicon, and language modeling that are trained separately. More recently, end-to-end (E2E) models that are trained to convert acoustic features to text transcriptions directly, potentially optimizing all parts for the end task \cite{graves2014towards}. 
With the performance of ASR systems reaching closer to that of a human, several efforts embarked to benchmark the performance of state-of-the-art ASR systems against professional transcribers \cite{xiong2016achieving,saon2017english}. In \cite{deepspeech2} the researches showed that E2E ASR can achieve competitive performance in simple speech recognition tasks like
reading newspaper. After that, a study by Microsoft \cite{xiong2016achieving}, suggested that the ASR systems have already reached the level of the professional human transcriber in more difficult ASR tasks like conversational speech. On the other hand, a study by IBM \cite{saon2017english} suggested that human parity in conversational  speech is still considerably better. The aforementioned studies were conducted on English language and were not explored in morphologically complex languages like Arabic. The Arabic language is the largest Semitic language with a high degree of affixations and derivations, which result in a huge increase in the number of word forms. Remarkably, the 400 million speakers (estimated in 2020) Arabic native speakers use Dialectal Arabic (DA) as their way of communication in the day-to-day speech. DA does not have standard orthographic rules. It can be argued that a language is a dialect with an army and navy \cite{michalowski2006lives}. If we take this perspective into consideration, we can describe the different Arabic dialects as different languages. However, Arabs in general perceive dialects as a deterioration from the classical Arabic, almost using all the same Arabic letters. An objective comparison of the varieties of Arabic dialects could potentially lead to the conclusion that Arabic dialects are historically related, but not synchronically, and are mutually unintelligible languages like English and Dutch. This makes Arabic an excellent choice to highlight the challenges of speech recognition in the wild.

Till today, the state-of-the-art ASR in the Arabic language comes from modular Hidden Markov Model Deep Neural Network (HMM-DNN) systems. Lately, the best ASR results on the Modern Standard Arabic (MSA) data were reported by the Aalto University team \cite{smit2017aalto}, with WER of 13.2\% on MGB2 test set using a combination of over 30 system. There are three major challenges when developing speech recognition models for the Arabic language:
\begin{itemize}
    \item Arabic is a consonantal language with most of the available text is non-diacritized. As a result, it is challenging to determine the location of the vowels, which can convey different meanings.
    \item Existence of different Arabic dialects with limited labeled data. Each dialect is a native Arabic language that is spoken, but not written, as it does not have standardized  orthographic  rules.
    \item Arabic morphological complexity with a high degree of affixation and derivation that makes it challenging to estimate probabilities for the language model and increases the out-of-vocabulary (OOV) rate.
\end{itemize}

\begin{figure}[hbt!]
\begin{center}
\includegraphics[width=11cm,height=9.5cm]{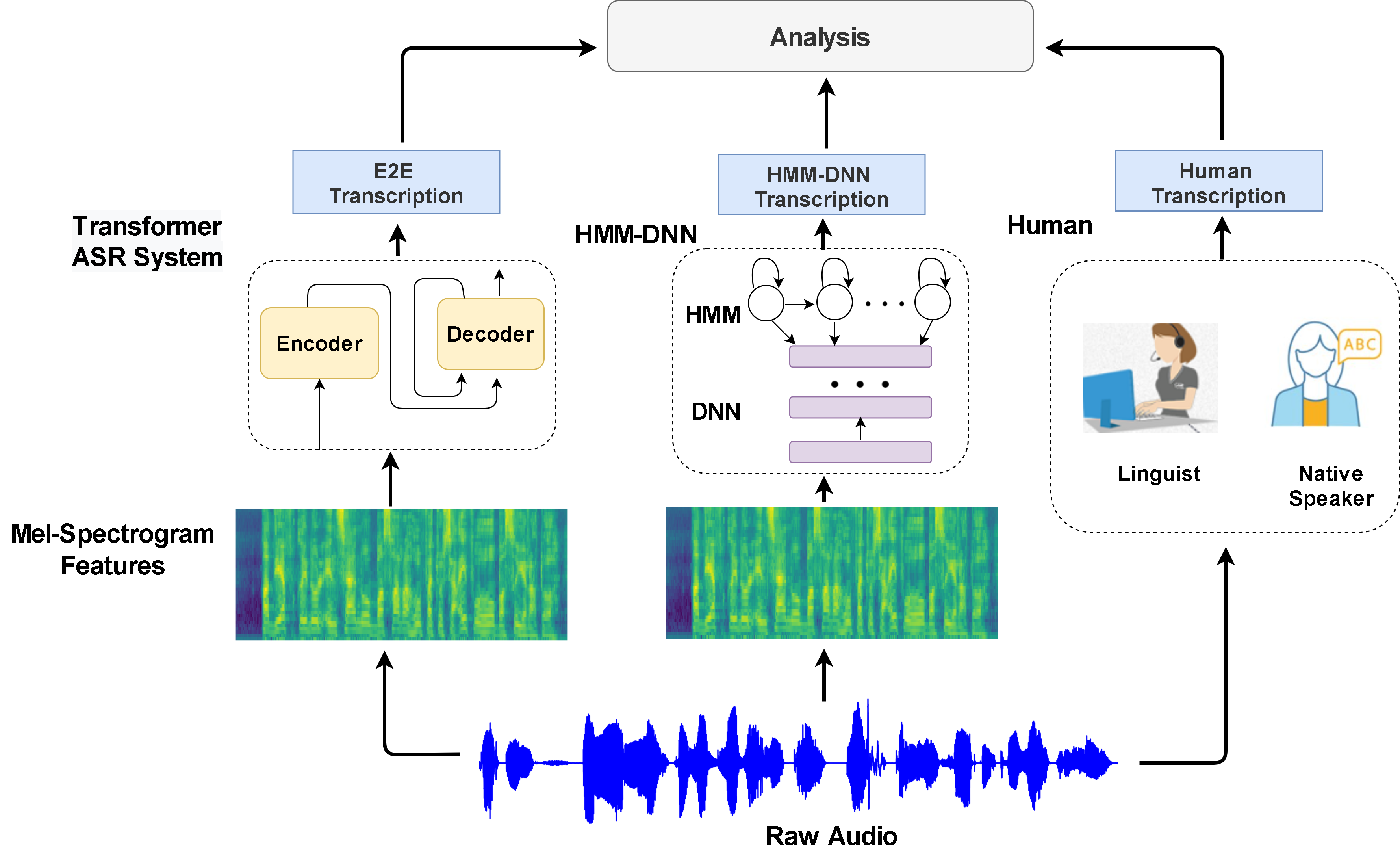}
\caption{ High-level illustration of the core study conducted in this paper. }
\label{high_level} 
\end{center}
\end{figure}

Several attempts have been made to address each of the aforementioned challenges. To address the nondiacritized words ambiguity, researchers in \cite{mubarak2019highly} utilized sequence-to-sequence deep learning model, inspired by Neural Machine Translation, to restore the missing diacritics. The proposed approach achieved new state-of-the-art with word error rate of 4.49\%. To address the challenge of limited dialect speech data, several transfer learning approaches were proposed \cite{das2015cross,khurana2019darts} that utilize similarity between MSA and dialectal Arabic (DA) speech. Finally, to deal with  morphological complexity and OOV, the character-level language model (LM) was suggested by \cite{ahmed2018end}. However, the major limitation with character level LM is that it is difficult to capture the contexts of the word in a sentence.
Furthermore, the previous approaches mainly used HMM-DNN models that have several limitations including model complexity (complicated to implement as HMM-DNN systems employ a modular design, with different modules for acoustic modeling, pronunciation lexicon, and language modeling, which are trained separately) and requirement of linguistic resources. 
The core contribution of this study lies in a new comprehensive analysis comparing the E2E transformer ASR, the modular HMM-DNN ASR, and the HSR. To avoid biases in our analysis, we collected a new evaluation set of 3 hours containing news reports and conversational speech with both MSA and DA. To better understand human performance, we hired expert linguists and educated native speakers to perform HSR task on the new hidden set. To the best of our knowledge, this is the first work that compares head to head single E2E, modular, and HSR performance. The main question that we address in this work is whether there are major qualitative differences between the HSR and the state-of-the-art machine results in the Arabic ASR as shown in Figure \ref{high_level}. In this work, we develop the first E2E transformer ASR for the Arabic language. Furthermore, we provide the best practices for finetuning the transformer ASR for dialectal Arabic\footnote{The source code has been made publicly available on Espnet Github repository https://github.com/espnet/espnet/tree/master/egs/mgb2/asr1}. 
 The proposed approach advances the Arabic ASR by addressing all the aforementioned HMM-DNN limitations, achieving significant improvements over the previous state-of-the-art in both MSA and DA tasks. The main advantage of E2E transformer architecture is that it reduces word ambiguity by learning the context with a self-attention mechanism. In order to address the OOV efficiently, we use sub-word tokenization \cite{kudo2018subword}. Furthermore, we benchmark the  performance of a single E2E transformer model with the state-of-the-art single HMM-DNN approach \cite{khurana2016qcri} pointing out the advantages and disadvantages of each approach. Finally, the proposed effectiveness of pretrained E2E transformer is evaluated on two dialectal datasets MGB3 and MGB5 using finetuning to highlight the challenges of Arabic ASR in the wild\footnote{In case of acceptance, the code will be 
made publicly available as part of Espnet official recipe}.

In summary, the key contributions of our work include:
\begin{itemize}
\item A comprehensive assessment of the human performance on Arabic speech recognition system, analyzing the type of errors and correlation with the machine.
   
\item A new milestone of the Arabic speech recognition performance with E2E transformer architecture for MSA and DA tasks.

\item As a part of HSR versus machine study, we provide a new hidden test set combining both MSA and DA to avoid being biased in our analysis, as previous test sets have been made public.
    
\end{itemize}
An additional contribution of our work is developing a voice activity detection (VAD) pipeline for the E2E transformer to address the problem of very long speech segments, which is a typical situation in a practical ASR setup. The developed pipeline combines the speech detection precision of InaSpeechSegmenter  \cite{ddoukhanmirex2018}, with the maximum length of the segmentation feature of energy-based VAD\cite{giannakopoulos2015pyaudioanalysis}. Finally, we plan to provide a new benchmark, including the state-of-the-art recipe, and pre-trained models and make them publicly available for the community.

The rest of the paper is organized as follows: In Section \ref{sec_related}, we cover related work. Section \ref{sec_models_descriptions} describes the developed ASR models in this study. Section \ref{sec_exp} describes the details of the experimental setup and the used datasets. Section \ref{sec_results} presents the results with comparison to the state-of-the-art approaches and human performance. Finally, Section \ref{sec_conclusion} concludes the findings of our study and discusses future work.  

\section{Related Work}\label{sec_related}
This section highlights prior work in ASR approaches, which can be grouped into modular HMM-DNN systems and E2E. In addition, we present a subsection about the studies that conducted comparisons between human and machine performance on speech transcription.

\subsection{Hybrid HMM-DNN systems}

For a long time, the HMM with Gaussian Mixture Model (GMM) has been considered the mainstream model for large vocabulary continuous speech recognition (LVCSR) achieving best recognition results. In \cite{dahl2011context}, authors proposed the first HMM-DNN hybrid approach, where GMM was replaced with the 
deep neural network model. It achieved significant performance gains compared to the HMM-GMM legacy system in the LVCSR task. After that, several DNN architectures were explored for the acoustic modeling, including Recurrent Neural Networks (RNNs), Bidirectional RNNs  (BDRNNs), and deep conditional random fields, which showed prominent performance improvement \cite{graves2013speech,hifny2015unified}.  In \cite{peddinti2015time}, the authors proposed a time delay neural network (TDNN) and showed better results in learning wider temporal dependencies compared to DNN and RNN based models. In \cite{khurana2016qcri}, researchers targeted addressing MSA task using similar architecture to \cite{peddinti2015time}, combining three AM models trained with Lattice-free Maximum Mutual Information (LF-MMI) objective function. They combined TDNN, LSTM, and BLSTM using the Minimum Bayes Risk (MBR) decoding criterion and achieved first place in the MGB2 Arabic Broadcast Media Recognition challenge with a WER of 14.2\%. After that, in MGB3 challenge \cite{ali2017speech} a team from Aalto University combined over 30 systems using MBR, including two acoustic models (TDNN-BLSTM) and a variety of language models (character, sub-word, and word-based) \cite{smit2017aalto}. Aalto team achieved a WER of 13.2\% on MGB2 and 37.5\% on MGB3, which is to the best of our knowledge, the state-of-the-art results. One main advantage of the modular HMM-DNN over the current E2E transformer system is that HMM-DNN is streamable while E2E-transformer is not. On the other hand, the main disadvantage of the modular HMM-DNN system is its complexity consisting of various components, including the acoustic model, language model, and pronunciation model. Each component of the system is optimized independently with a different objective function, which usually leads to a non optimal global solution. In addition, HMM-DNN systems require linguistic resources, such as handcrafted pronunciation dictionary that is subject to human error.

\subsection{End to end ASR systems}

E2E deep learning models were introduced to simplify the complexity of HMM-DNN modular models into a single deep network architecture and address the aforementioned limitations. The main issue with E2E sequence to sequence models was data alignment \cite{wang2019overview}. Several approaches were introduced to address the data alignment problem, including connectionist temporal classification (CTC) based model \cite{graves2014towards} and attention-based  sequence to sequence model \cite{chan2016listen,chorowski2015attention}. The E2E CTC approach uses Markov assumptions to efficiently solve sequential problems with dynamic programming. However, the main limitation of CTC is the assumption that the label outputs are conditionally independent of each other. On the other hand, the attention-based E2E model uses an attention mechanism to perform alignment between acoustic frames and label sequence without the independence assumption, yet it can result in nonsequential alignments.  In \cite{watanabe2017hybrid}, the authors proposed a multitask CTC/attention approach, which effectively utilizes the advantages of both architectures in training and decoding. The proposed multitask CTC/attention approach is based on long short term memory (LSTM) sequence to sequence model. At this point it is  important to note that CTC and attention losses are only surrogates for the actual performance measure which is word error rate, while HMM-DNN systems are trained with objective functions which are close approximations to the word error rate ( MMI or MBR). In addition, attention and CTC objectives are teacher-forcing based likelihood and do not consider beam search with an LM rescoring. On the other hand, MMI and MBR objectives try to optimize ASR beam search decoding (through lattice, n-best in the classical one or lattice-free MMI in the chain model).
In the literature, we found only two works that are related to E2E models  for Arabic speech data \cite{ahmed2018end, belinkov122019analyzing}. In \cite{ahmed2018end} the authors proposed an E2E model based on Bidirectional Recurrent Neural Networks (BRNN) with CTC. However, the results in \cite{ahmed2018end}  were only reported on the MGB2 developments set with no results on the test set. In \cite{belinkov122019analyzing}, the researchers studied the learned internal representations in DeepSpeech2 E2E  \cite{deepspeech2} on phonemes and graphemes classification tasks.
The transformer-based architecture was first introduced as a neural machine translation system \cite{vaswani2017attention} to replace the recurrence with the self-attention mechanism. In \cite{karita2019comparative,wang2020transformer}, authors showed superiority performance of transformer-based models compared to state-of-the-art recurrent networks in ASR task. Besides, the E2E transformer-based ASR showed close performance to the state-of-the-art HMM-DNN systems \cite{synnaeve2020end}.
In this paper, we develop an E2E transformer-based ASR model with a multitask CTC/attention objective function for Arabic speech data.  

\subsection{Speech recognition by humans and machines}
Advances in ASR technology produced major improvements reaching the range of human performance. In two papers by Microsoft \cite{xiong2016achieving,stolcke2017comparing}, authors suggested that the machine has already reached the human performance in English ASR. Microsoft reported that their improved ASR outperformed expert transcriber by 0.1\% and 0.2\% WER on Switchboard and CallHome datasets, respectively. In their study, an existing Microsoft transcription pipeline was leveraged, in which transcription is conducted on a weekly basis. The transcription was conducted on NIST 2000 test set with two passes. In the first pass, a transcriber works from scratch to transcribe the data and, in the second pass, a second listener monitors the data to do error correction. On the other hand, IBM \cite{saon2017english} conducted an independent set of human performance measurements to verify the aforementioned claims by Microsoft. IBM found that human performance is considerably better outperforming their state-of-the-art ASR by 0.4\% and 3.5\% WER on Switchboard and CallHome datasets, respectively. Unlike the previous Microsoft setup, IBM experiment transcribers were aware of the experiment and were actively involved. Three independent transcribers were used for the first pass in addition to quality control by a fourth senior transcriber. The final performance was chosen based on the lowest transcriber word error rate (WER). Unlike previous studies where several systems were combined, in this paper, we are looking for a practical ASR usage where our focus is to get the best results from a single ASR system. Furthermore, in the present study, best practices of the two aforementioned studies are followed to set the most realistic and unbiased comparison between human and machine as described in Section \ref{human_trans}.

\section{ASR Models Description}\label{sec_models_descriptions}

\subsection{Hybrid HMM-DNN}

The acoustic model for the modular system is trained using the 1,200 hours MGB2 \cite{ali2016mgb}. For language modeling, we use the 130M words crawled from the Aljazeera Arabic website from the period 2000 - 2011, as provided for the MGB2 challenge. LM experiments used a grapheme lexicon of 1.3M words. The grapheme-based lexicon has a 1:1 word-to-grapheme mapping, which means the vocabulary size is the same as the lexicon size.  

\noindent \textbf{Acoustic modeling}: In this study, we adopt the architecture proposed by \cite{peddinti2015time}, which consists of combining a time delay neural network (TDNN) with long short term memory (LSTM) layers and showed significantly better results compared to bidirectional LSTM (BLSTM) acoustic modeling. The TDNN-LSTM model consists of 5 hidden layers, each layer containing 1,024 hidden units. Neural networks are trained using lattice-free maximum mutual information (LF-MMI) \cite{povey2016purely}. The input to the modular system is standard 13-dimensional cepstral mean-variance normalized (CMVN) Mel-Frequency Cepstral Coefficients (MFCC) features without energy, and its first and second derivatives. For each frame, we also include its neighboring $\pm4$ frames and apply Linear Discriminative Analysis (LDA) transformation to project the concatenated frames to 40 dimensions, followed by Maximum Likelihood Linear Transform. Speaker adaptation is also applied with feature-space Maximum Likelihood Linear Regression (fMLLR). The GMM-HMM model has 100K Gaussians for 5K states. The parameters for language weight and silence penalty that provided the best results on the development set were found 0.8 and 0.0 respectively. Acoustic models are built using Kaldi ASR toolkit \cite{povey2011kaldi}.\\
\textbf{Language modeling}: Two \textit{n}-gram LMs are trained: a big four-gram LM (bLM$4$), trained using the spoken transcripts and the 130M words background text; and a smaller four-gram LM obtained by pruning bLM$4$ using pocolm\footnote{https://github.com/danpovey/pocolm}. The small LM is used for first-pass acoustic decoding to generate lattices. These lattices are then rescored using the bLM$4$.
In an attempt to be consistent with the E2E system, we explored using the time-restricted self-attention layer for the acoustic modeling as described in \cite{povey2018time}. We also explored the sub-word modeling using 1K sub-words as it has been tuned for the HMM-DNN. Our results showed 1.1\% relative reduction in WER on the MGB2 development set and 0.7\% relative increase in WER on the MGB2 test set compared to word tokenization. As a result, we decided to drop the sub-word modeling and adopt the state-of-the-art TDNN-LSTM architecture with word tokenization. Throughout the paper we refer to this modular system as HMM-DNN.

\subsection{E2E transformer}
\begin{figure}[hbt!]
\begin{center}
\includegraphics[width=10cm,height=10cm]{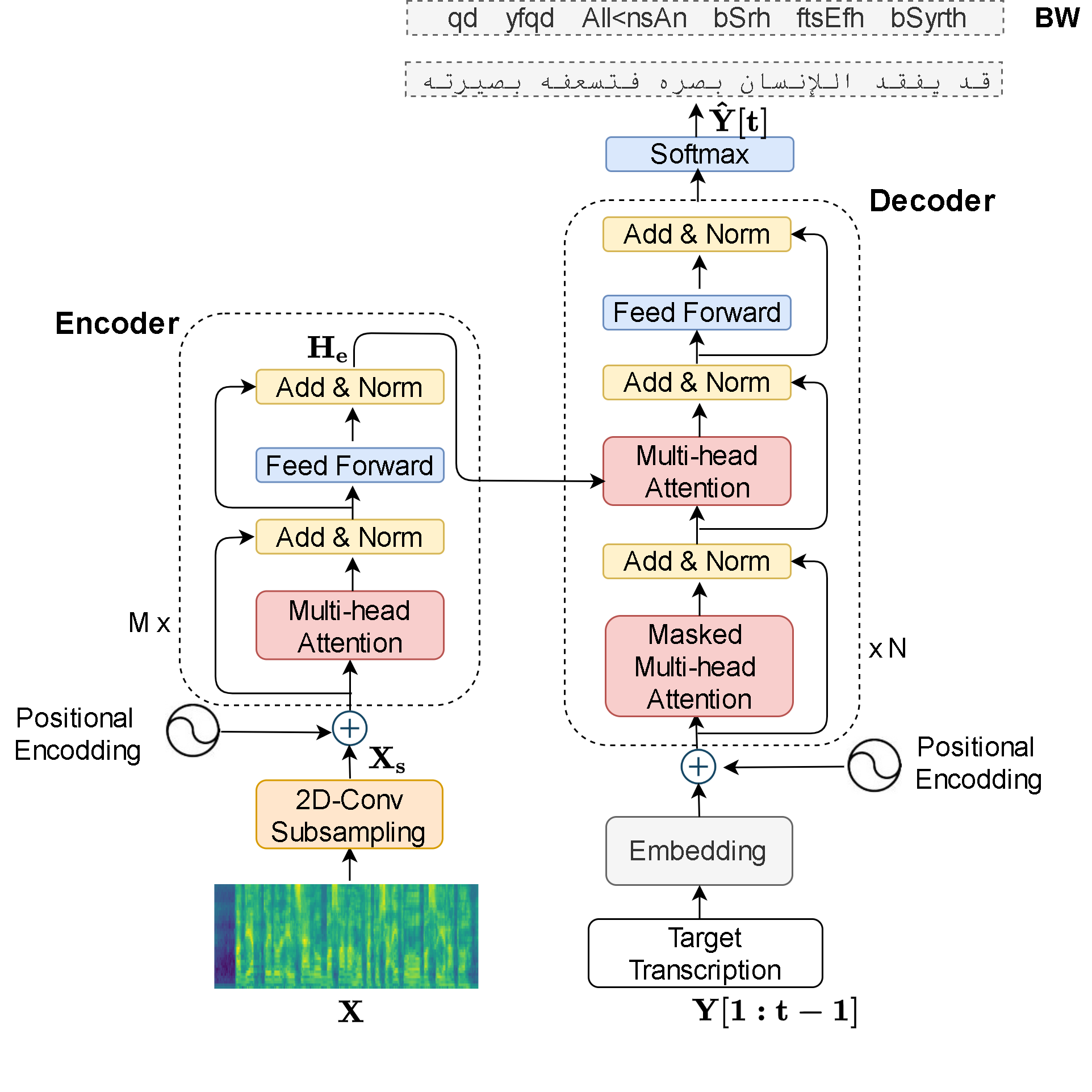}
\caption{Illustration of E2E transformer-based ASR architecture. The input to the encoder during the training is a sequence of feature frames. The output of the encoder is fed as input to the decoder in addition to the masked target transcription. The output of the decoder is the prediction of the masked transcription}
\label{transformer_diagram} 
\end{center}
\end{figure}

Transformer is a recent sequence to sequence model that completely 
replaced the recurrence in traditional recurrent networks with  self-attention mechanism and sinusoidal position information. In this paper, we utilize transformer-based architecture for Arabic ASR, as shown in Figure \ref{transformer_diagram}. On a very high level, the transformer consists of an encoder model with $M$ repeating encoder blocks and a decoder model with $N$ repeating decoder blocks. The encoder model mainly maps the input vector to a latent representation. The input to the encoder $\mathbf{X}$ is a sequence of 83-dimensional feature frames, 80-dimensional log Mel spectrogram with pitch features \cite{ghahremani2014pitch}. The decoder generates one prediction at a time in an auto-regressive fashion. At each time step, the input to the decoder model is the latent representation for the encoder model and previous decoder predictions. 

\subsection{Pre-encoder}
First, the acoustic feature frames $\mathbf{X}$ are transformed into sub-sampled sequence $\mathbf{X_{s}}$ $\in$ $\mathbb{R}^{d^{sub}\times d^{\mathrm{model}}}$ with 2D-CNN sampling layer. The $d^{sub}$ is the length of the output sequence and $d^{model}$ is the number of input feature dimensions to the Encoder.

\subsection{Encoder}
Each encoder block consists of sub-layers: a multi-head self-attention mechanism and a position-wise fully connected network.
The output of each sub-layer is followed by a layer normalization \cite{ba2016layer} with a residual connection from the sub-layer input \cite{he2016deep}. The input to the first encoder block is the sub-sampled sequence input $\mathbf{X_{s}}$. At the self-attention sub-layer, the $\mathbf{X_{s}}$ is transformed to queries $\mathbf{Q} = \mathbf{X_{s}} * \mathbf{W^q}$, keys $\mathbf{K} = \mathbf{X_{s}} * \mathbf{W^{k}}$, and values $\mathbf{V} = \mathbf{X_{s}} * \mathbf{W^{v}}$,
where $\mathbf{W^{q}}$ and $\mathbf{W^{k}}$ $\in$  $\mathbb{R}^{d^{model}\times
d^{\mathrm{k}}}$ and $\mathbf{W^{v}}$ $\in$ 
$\mathbb{R}^{d^{model}\times  
d^{\mathrm{v}}}$ are learnable weights. 
The $d^{model}$ is the output dimension from the previous attention layer, 
and $d^{\mathrm{v}}$, $d^{\mathrm{k}} = d^{\mathrm{q}}$ are the dimensions of values, keys and queries. After that, a normalized weighted similarity $Z$ from self-attention is obtained with a softmax as shown in Eq \ref{eq1}.
\begin{equation}\label{eq1}
SelfAttention(\mathbf{Q,K,V}) = softmax(\dfrac{\mathbf{Q} * \mathbf{K}^T}{\sqrt{d^k}})*\mathbf{V} 
\end{equation}
To deal with multiple attentions, multiple attention sub-layers are used in parallel, usually referred to as multi-head attention (MHA). The MHA is obtained by concatenating all of the self-attention heads at a particular layer.

\begin{equation}\begin{aligned}
\operatorname{MHA}(\mathbf{Q,K,V}) &=\left[\mathbf{Z_{1}, Z_{2}}, \cdots, \mathbf{Z}_h\right] \mathbf{W}^{\mathrm{h}} \\
\mathbf{Z}_{i} &=\operatorname{SelfAttention}( \mathbf{Q}_{i}, \mathbf{K}_{i},  \mathbf{V}_{i})
\end{aligned}
\end{equation}
where $h$ is the number of attention heads in a layer and $i$ corresponds to the $i^{th}$ head in the layer.
The output from MHA is normalized and then fed to the Feed Forward (FF) sub-layer connected network which is applied to each position separately.

\begin{equation}\operatorname{FF}(\mathbf{z}[t])=\max \left(0, \mathbf{z}[t]* \mathbf{W}_{1}+\mathbf{b}_{1}\right) \mathbf{W}_{2}+\mathbf{b}_{2}\end{equation}
where $\mathbf{z[t]}$ is the $t^{th}$ position of the input $\mathbf{Z}$ to the FF sub-layer. 

\subsection{Decoder}
The decoder architecture is very similar to the encoder. However, in addition to MHA self-attention and fully connected sub-layers, it has a third masked self-attention layer. The masked self-attention in the decoder is allowing to attend only to earlier positions in the output sequence. The decoder prediction $\hat{Y}[t]$ at each time step is conditionally dependent on the final representation produced by the encoder $\mathbf{H}_e$ and the previous target sequence $Y[1:t-1]$. The conditional dependence is obtained with the multihead attention that calculates the attention between the encoder latent features and the previous decoded sequence. Similar to the encoder, the decoder has residual connections and layer normalization around each sub-layer.

\subsection{Positional encoding}
Positional encoding is added to the input embeddings to reflect the positional context of each word in the sentence. Transformers use sinusoidal positional encoding with different frequencies as shown in Equation \ref{eq2}.

\begin{equation}\label{eq2}
\begin{aligned}
P E_{(n, 2 i)} &=\sin \left(n / 10000^{2 i / d_{\text {model }}}\right) \\
P E_{(n, 2 i+1)} &=\cos \left(n / 10000^{2 i / d_{\text {model}}}\right)
\end{aligned}\end{equation}
where $n$ is the position of a word in the sentence and $i$ is a position along the embedding vector dimension.

\subsection{Transformer ASR training}
During training, the acoustic model predicts the posterior probability of the transcription $\mathbf{Y}$ given acoustic features $\mathbf{X}$. The total objective function of the acoustic model  $\mathcal{L}_{asr}$ is a multi-task learning objective that combines the E2E decoder loss  $\mathcal{L}_{\mathrm{d}} = - log[P_d(\mathbf{Y|X})]$ and the CTC loss $\mathcal{L}_{\mathrm{ctc}} = - log[P_{ctc}(\mathbf{Y|X})]$. 
This multi-objective function was proposed to improve model robustness and faster convergence  \cite{watanabe2017hybrid}.

\begin{equation}\label{eq3}
\mathcal{L}_{asr}=\alpha \mathcal{L}_{\mathrm{ctc}}+(1-\alpha) \mathcal{L}_{\mathrm{d}}
\end{equation}
where $P_d$ are the probabilities predicted by transformer decoder, $P_{ctc}$ are the CTC probabilities, and $\alpha$ is a weighting factor that trades-off the two losses. 

\subsection{Transformer ASR inference}

During the inference mode, a language model (LM) is used to disambiguate between hypothesised words generated by the decoder. In particular, two types of LM models are used in this paper: long short term memory (LSTM) and  transformer-based language model (TLM). The LM  model prediction is combined with E2E as  shown in Eq. \ref{eq4}. 
\begin{equation}\label{eq4}
\hat{Y}= \argmaxA_{Y\in Y^*} \{\lambda \mathcal{L}_{\mathrm{ctc}}+(1-\lambda) \mathcal{L}_{\mathrm{d}} + \mu\mathcal{L}_{lm}\}
\end{equation}
where $\mathcal{L}_{lm} = P_{lm}(\mathbf{Y})$ is the language model prediction, $\mathbf{Y}^*$ is a set of hypotheses generated using beam search over all possible sequences, $\mu$ and $\lambda$ are trade-off factors. 
The audio segmentation during the inference in an experimental setup is usually assumed to be prepared by an expert transcriber. In Section \ref{VAD}, we study the effect of segment duration variability on E2E transformer performance.

\section{Experimental Setup} \label{sec_exp}
The proposed E2E transformer approach is  benchmarked with state-of-the-art approaches on MSA task with (MGB2) data and on dialectal task with (MGB3, MGB5) data. 
For MSA evaluation, the conventional word error rate (WER) is used. However, for dialectal data evaluation, the  multi-reference word error rate (MR-WER) and averaged WER (AV-WER) are adopted from MGB3, MGB5 challenges \cite{ali2017speech,ali2019mgb}. The MR-WER was proposed to evaluate dialectal data, which does not have standardized orthography \cite{ali2015multi}. All models are implemented  using  Espnet toolkit \cite{watanabe2018espnet}. We ran our experiments on an HPC node equipped with 4 NVIDIA Tesla V100 GPUs with 16 GB memory, and 20 cores of Xeon(R) E5-2690 CPUs. 

\subsection{Human transcription experiment}\label{human_trans}

Two professional linguist transcribers were hired independently and their quality transcription was checked by a third senior transcriber with extensive experience in the linguistic annotation. In addition, another three educated native
Arabic speakers (not linguists) were hired to transcribe the same data. The transcribers were not aware of the experiment conducted and they performed the transcription as a part of their daily transcription tasks. The same guidelines were provided to all transcribers to ensure the consistency and the quality of the transcription, which included guides for audio segmentation, truncated words, words from other languages, hesitation, etc. There were no restrictions on the number of times to listen to the speech. We found that, on average, transcribers needed to listen between 2-4 times for each sentence. 
As part of our study, $3$ hours of Arabic speech data from Aljazzera news channel in the period from (June - August) 2020 was collected. The data included a variety of conversations, interview programs and reports by journalists from the field. Around 10\% of the data is in Dialect Arabic, including Egyptian, Gulf, Levantine, and North African.
This dataset is also used  as a final hidden test (hereafter referred to as  Hidden\_Test\footnote{As part of our ongoing effort, this dataset is hosted on https://arabicspeech.org/ as the final hidden test set for Arabic ASR benchmarking.}) as it was not seen by any model before.

\subsection{Model development data}
In this work, the Arabic Multi-Genre Broadcast (MGB2) corpus \cite{ali2016mgb} was used for model training. 
Around 70\% of the data is considered Modern Standard Arabic (MSA), with the rest in Dialectal Arabic including Egyptian (EGY), Gulf (GLF), Levantine (LEV), and North African (NOR). The dataset span is more than 10 years recording, during 2005-2015, from 19 distinct programs, and contains around 1,200 hours. The programs include conversations (63\%), interviews (19\%), and reports (18\%). The conversational speech is the most challenging because it includes overlapping speech with multiple dialects. All programs were aligned using the QCRI Arabic LVCSR system \cite{ali2014complete}, which is grapheme-based with one unique grapheme sequence per word. Moreover, the dataset includes a large corpus of background text that can be used to build a language model. The text corpus consists of over 130 million words crawled from the Aljazeera.net website. BuckWalter\footnote{Buckwalter (BW) is a one-to-one mapping allowing non-Arabic speakers to understand Arabic scripts, and it is also left-to-right, making it easy to render on most devices.} (BW)
mapping format is used for the transcriptions and the background text data. More details about the data can be found in Table \ref{mgb2_1}.

\begin{table}[!ht]
  \begin{center}
    \caption{Arabic datasets description used in this study.}
    \label{mgb2_1}
    \resizebox{0.8\textwidth}{!}{\begin{tabular}{lrrrr}
      \toprule %
      \textbf{Dataset} & \textbf{Type} & \textbf{Hours} & \textbf{Programs} & \textbf{\#Segments}     \\
      \midrule
      & Training & 1200  & 2,214 & 370K   \\ 
      \textbf{MGB2} & Development & 10  & 17 & 5002 \\
        & Evaluation & 10  & 17 & 5365 \\
        \midrule
         & Adaptation & 4.6  & 23 & 2202 \\
        \textbf{MGB3} & Development & 4.8  & 24 & 2181 \\
         & Evaluation & 6  & 30 & 5746 \\
        \midrule
        & Adaptation & 10.2  & 69 & 31063 \\
        \textbf{MGB5} & Development & 1.3  & 10 & 1129 \\
        & Evaluation & 1.4  & 14 & 1055 \\
        \midrule
        \textbf{Hidden\_Test} & Evaluation & 3  & 7 & 1404 \\
        
      \bottomrule
      
    \end{tabular}}
  \end{center}
\end{table}

 
    
      
       
       

    
      
       

\subsection{Dialectal data}\label{Hidden_Test}
In this study, the proposed E2E transformer ASR is benchmarked on two dialectal real-world datasets; the Egyptian MGB3 \cite{ali2017speech} and the Moroccan MGB5 \cite{ali2019mgb}. The MGB3 dataset comprises of 16 hours of speech obtained from 80 YouTube videos, while MGB5 consists of 13 hours of speech extracted from 93 YouTube videos. Both datasets are distributed across seven genres: comedy, cooking, family/kids, fashion, drama, sports, and science talks (TEDx).

\subsection{Data pre-processing}\label{Data_prep}
The raw audio segments were first augmented with the speed perturbation approach, which increased the original signal by a factor of three with speed factors of 0.9, 1.0 and 1.1 \cite{ko2015audio}. Each augmented audio was transformed to a sequence of 83-dimensional feature frames for the E2E model, and an 80-dimensional log Mel spectrogram with pitch features \cite{ghahremani2014pitch}. In addition, the resulting mel-spectrogram features were augmented with specaugment approach \cite{park2019specaugment}, which consists of three deformations: warping the features in time direction, masking blocks of consecutive frequency channels and masking blocks of utterances in time.
As for the text data for the language model development, two sources were considered: 
the transcription text and the background text of 130 million words \footnote{https://www.aljazeera.net/}. The data was cleaned by removing punctuations, diacritics, extra empty spaces, newlines, and single-character words. To overcome the problem of very long sequences, the text was segmented to contain a maximum of 200 words with an overlap of 50 words \cite{pappagari2019hierarchical}.
The sub-word model \cite{kudo2018subword} was used to tokenize the input text and prepare the vocabulary.

\subsection{Default Model Hyperparameters}\label{hyperparameter_tuning}

All hyperparameters were obtained using a grid search. The parameters tuning was performed on a small subset of MGB2 data (250 h). The E2E transformer-based ASR model was trained using Noam optimizer \cite{vaswani2017attention} with a learning rate of 5. The best values for  multi-objective tradeoff weights: $\alpha$ in Equation \ref{eq3}, $\mu$ and $\lambda$ in Equation \ref{eq4} were found to be 0.3, 0.3 and 0.5, respectively.
Table (\ref{hyp_ASR_trans}) summarizes the best set of parameters that were found for AM and LM transformer architecture\footnote{The source code to reproduce the results is made publicly available on Espnet Github repository https://github.com/espnet/espnet/tree/master/egs/mgb2/asr1}. As for LSTM LM, the best results were obtained with 2 layers and 650 units/layer. The LSTM LM was trained with a batch-size of 512 and a stochastic gradient descent algorithm with a learning rate of 1. The perplexity for E2E-Transformer LM and LSTM LM on the MGB2 entire text data (background text + transcription) after 10 epochs are 36.5 and 54.5 respectively.

\begin{table}[!ht]
\centering
\caption{Values of tuned hyperparameters for E2E AM transformer and LM transformer obtained from grid search.}
\label{hyp_ASR_trans}
\resizebox{1\textwidth}{!}{
\begin{tabular}{|c|c|c|}
\hline
 {\textbf{}}& \textbf{AM Hyperparameters} & \textbf{LM Hyperparameters}\\\hline
  Input & batch-bins: $22000000$ & batch-size: $64$
  \\ \hline
 Encoder& $12$ layers, $8$ attention heads$/$layer & $12$ layers, $4$ attention heads$/$layer
\\ \hline
Decoder & $6$ layers, $8$ attention heads/layer & $12$ layers, $4$ attention heads/layer

\\ \hline
$d^{model}$ (attention) & $512$ & $512$

\\ \hline

FFN & $2048$ & $2048$\\ 
\hline
\end{tabular}}
\end{table}

\section{Results and Discussion}\label{sec_results} 

\subsection{ASR benchmarking}
The developed E2E Transformer (E2E-T) is benchmarked with the state-of-the-art modular system; \cite{khurana2016qcri}, Aalto system \cite{smit2017aalto} and the expert linguist. In addition, the contribution of CTC and Attention objectives to the E2E-T overall performance is examined. E2E-T(CTC+Attention) hereafter referred to as just E2E-T. Table \ref{soa_results} summarizes the WER results on the MSA datasets: the MGB2 test set and the Hidden\_Test set. It worth nothing that the WER results were scored with a global map scoring (GLM) file to map the numbers from digits form to their verbatim transcription.
\begin{table}[!ht]
  \begin{center}
    \caption{WER\% performance of E2E transformer (E2E-T) with (CTC, Attention and CTC+Attention) , HMM-DNN and state-of-the-art Aalto approach. The results were obtained on MGB2 test set and the Hidden\_Test set that was collected as part of this study.}
    \label{soa_results}
    
   \resizebox{1\textwidth}{!}{ \begin{tabular}{|c|c|c|c|c|c|}
      \toprule 
       & \textbf{HMM-DNN} & \textbf{Aalto} \cite{smit2017aalto} &\textbf{E2E-T(CTC)} & \textbf{E2E-T(Att)}  & \textbf{E2E-T(CTC+Att)}\\ 
       \midrule
      \textbf{MGB2\_Test} & 15.8  & 13.2 & 16.9  & 13.4  & \textbf{12.5}  \\
       \textbf{Hidden\_Test} & 15.9  & -  &  17.7 & 14.7  & \textbf{12.6} \\
     
      \bottomrule
    \end{tabular}}
  \end{center}
\end{table}
It can be seen that the proposed E2E Transformer with hybrid (CTC+Attention) outperforms the single HMM-DNN and Aalto system by 20\% and 5\% in relative WER, respectively. In addition, it can be seen that most of the contribution to the overall E2E-T performance is coming from the attention objective achieving WER of 13.4\% and 14.7\% on MGB2\_Test and Hidden\_Test sets respectively. Adding CTC improves WER by an absolute of 0.9\% and 2.1\% on MGB2\_Test and Hidden\_Test sets respectively.

\subsection{Human and machine error analysis}
\begin{figure}[!ht]
      \includegraphics[width=10cm,height=9cm]{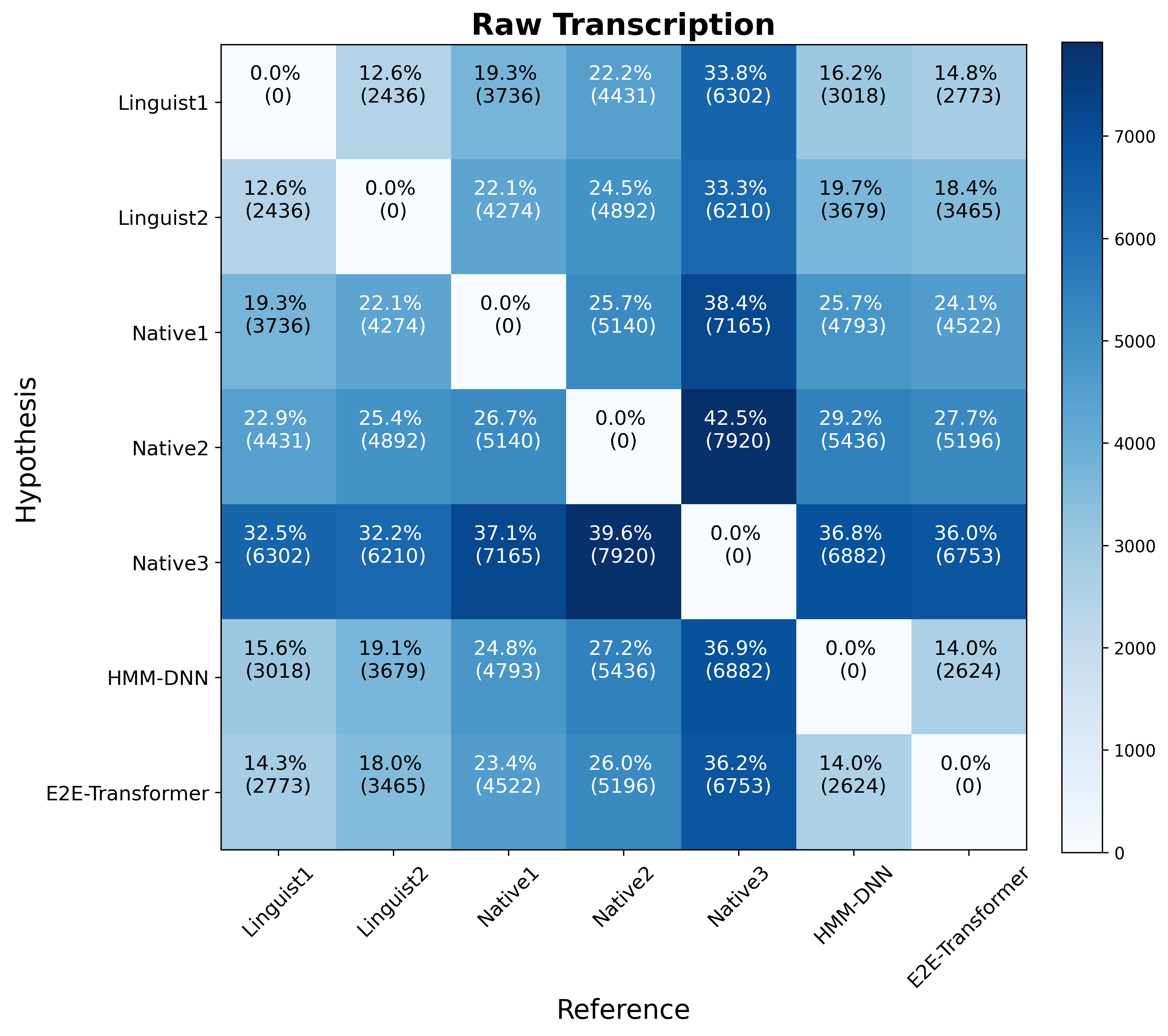}
    \caption{Confusion matrix of inter-annotation disagreement from raw transcription text. The rows represent a hypothesis and columns represent corresponding reference.}
    \label{conf0}
\end{figure}

\begin{figure}[!htb]
      \includegraphics[width=10cm,height=9cm]{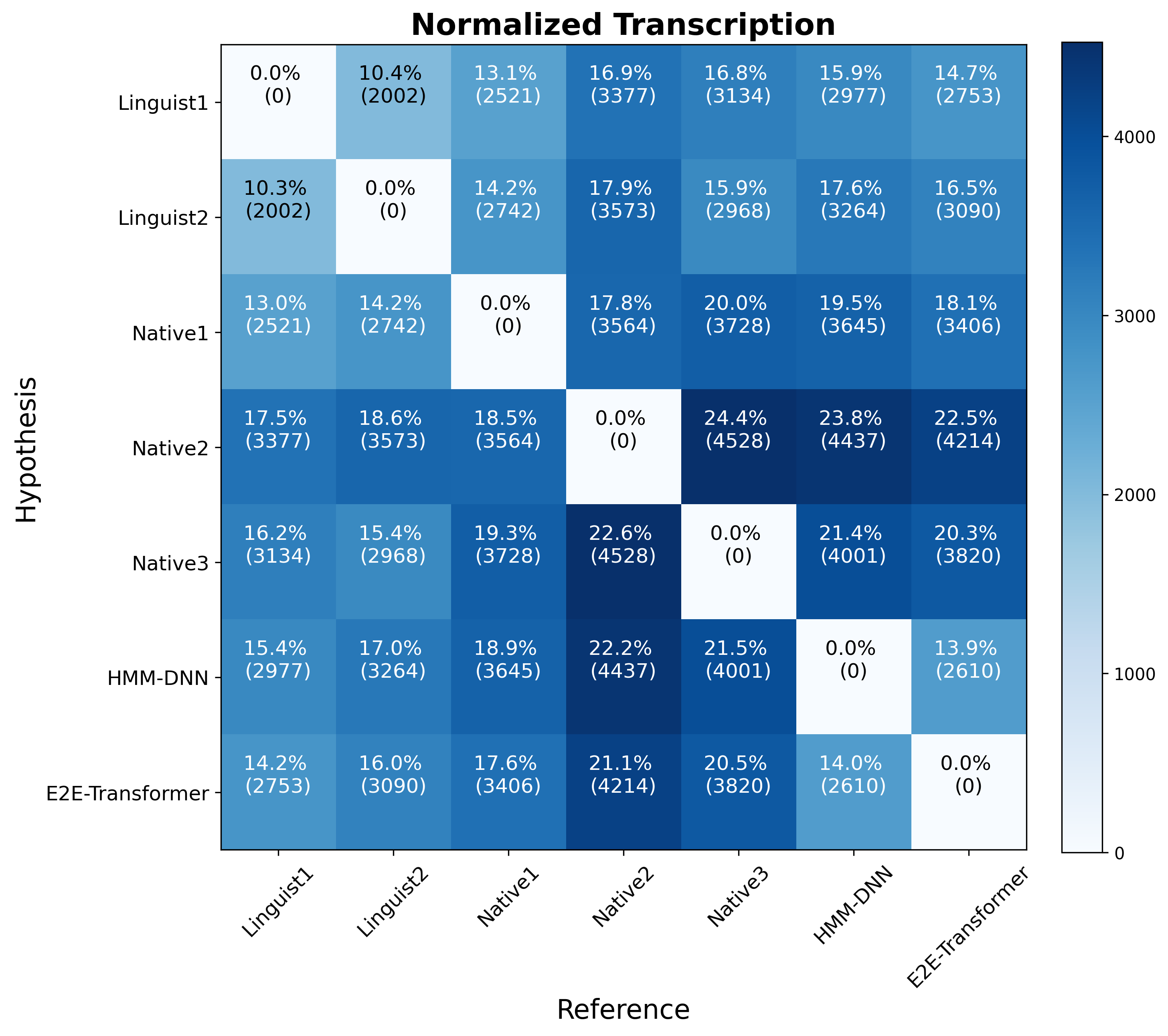}
    \caption{Confusion matrix of inter-annotation disagreement from normalized transcription text.}
    \label{conf1}
\end{figure}

In this section, we analyze in more details the type of errors and correlation between the expert linguist (Linguist), native speaker (Native), E2E transformer and HMM-DNN ASR systems. Figures (\ref{conf0},\ref{conf1}) illustrate the inter-annotation disagreement on the Hidden\_Test data for raw and normalized text. It can be seen from Figures (\ref{conf0},\ref{conf1}) that the inter-annotation disagreement between the two linguists is between 10.4\% and 12.6\% for raw and normalized transcriptions respectively.
We found that the inter-annotation disagreement between the two linguists is mainly caused by the variability in transcribing dialectal speech, which does not have orthography standards.  In addition, we observe that the expert linguists tend to pay more attention than the native speaker to linguistic mistakes especially with Alif/Ya/Ta-Marbuta, which are common mistakes in the Arabic language. As a result we reduced the disagreement with the common Alif/Ya/Ta-Marbuta normalization. The normalization removes distinctions within three sets of characters that are often written inconsistently in DA and sometimes in MSA: Alif forms ( A = \<ا> , $>$ = \<أ>,
$<$ = \<إ>, $|$ = \<آ>) , Ya forms ( y = \<ي> , Y = \<ى>, and Ta-Marbuta forms ( p = \<ة> , h = \<ه>).
Comparing Figures (\ref{conf0} \ref{conf1}) it can be seen that the inter-annotation disagreement between the linguist and the native speaker was significantly reduced by up to 18\%  absolute, which shows that almost half of the disagreement between the linguist, and the Native is due to linguistic mistakes. 
For the comparison of inter-annotation disagreement with a group, we define the inter-annotation disagreement gap $G()$ (hereafter referred to as gap) between $a_i$ a member of group $A$ and a group $B$ mathematically as $G(a_i,B)=\frac{1}{J+K} \mathop{\sum_{j=1}^{J} \sum_{k=1}^{K}} abs(disag(a_i,b_j) - disag(b_j,b_k))$ $\forall j \neq k$, where $disag()$ is the inter-annotation disagreement.  
Table \ref{int-gap} summarizes the inter-annotation disagreement gaps of the best machine (E2E-Transformer) with the Native speaker group and the expert Linguist group.
\begin{table}[!ht]
  \begin{center}
    \caption{The inter-annotation disagreement gaps from raw and normalized transcriptions. The gaps are computed between the E2E-Transformer and Native speaker group, E2E-Transformer and expert Linguist group, Native and expert Linguist group.}
    \label{int-gap}
    \resizebox{0.6\textwidth}{!}{%
    \begin{tabular}{|l|c|c|c|c|}
      \toprule 
        &
       \multicolumn{2}{c|}{\textbf{Native}} & \multicolumn{2}{c|}{\textbf{Linguist}} \\
      \cline{2-5}
       & Raw & Norm & Raw & Norm
      \\
       \midrule
       E2E-Transformer & 6.6 & 2 & 3.5 & 4.7  \\
       \midrule 
       Native & - & - & 13.1 & 5.4 \\
      \bottomrule
    \end{tabular}}
  \end{center}
\end{table}
One can observe that the inter-annotation gap of disagreement between the E2E-Transformer ASR and the expert linguist is 3.5\%, 4.7 \% from the raw and normalized transcriptions respectively. Hence the minimum gap for the machine to overcome the expert linguist performance is 3.5\%. It noticeable that the gap between the native speaker and the expert linguist compared to the gap between the E2E-Transformer and expert linguist is higher by 9.6\%, 0.7 \% for the raw and normalized transcriptions respectively.\\
Furthermore, we take a closer look at the most frequent top ten errors in terms of substitutions, deletions, and insertions made by the linguist and E2E transformer, as shown in Tables \ref{substitusions} and \ref{insertion_deletion}. 
\begin{table}[!ht]
  
    \caption{Most common substitutions for E2E ASR system and expert linguist. The number of times each error occurs is followed by the word in the
reference and the corresponding hypothesis.}
    \label{substitusions}
    \centering
    \resizebox{\textwidth}{!}{%
    \begin{tabular}{|c|c|c|c|c|}
      \toprule 
      \multicolumn{2}{|c|}{\textbf{E2E Transformer}} &  \multicolumn{2}{c|}{\textbf{Linguist}} \\ 
     \hline
      BW (REF / HYP)& REF Translation &  BW (REF / HYP) & REF Translation\\ 
      \hline
      \colorbox{green}{12: $<$nh / $>$nh} & it is  &  \colorbox{green}{11: $<$nh / $>$nh} & it is  \\
       12: AldEwY / AldEwp & lawsuit &  \colorbox{green}{6: btAEh / btAEt} & possessed  \\ 
       7: AljAbry / Aljbr & Aljabry (name) & 7: lk / lky & your \\ 
      6: dyAb / AldyAb & Dyab (name) & \colorbox{green}{6: nHnA / nHn} & we \\
     \colorbox{green}{5: btAEh / btAEt} & possessed  &  4: tfDl / AtfDl & go ahead \\
       \colorbox{green}{5: nHnA / nHn} & we  &  4: fy / fyh & in \\
      4: Ally / Alty & which & \colorbox{green}{4: l$>$n / $>$n} & to\\
     4: AlwA\$nTn / wA\$nTn & Washington & 3: AlmAyh / AlmwyA & water \\
       \colorbox{green}{ 4: l$>$n / l\}n} & to & 3: ln / lm  & did/will not\\
       3: lAzAlt / zAlt& still & 3: bdt / bd$>$t & started \\
       
      \bottomrule
    \end{tabular}}
\end{table}
Similar errors from both the E2E transformer and the linguist are highlighted with the same color. Inspections revealed that the top errors made by both human and machine are substantially similar, especially the insertions and deletions. Looking at the substitutions, one can notice that most of the errors for both the linguist and E2E are on dialectal words. The main difference between E2E and the linguist is that the E2E tends to make more errors on rare Arabic words like names. In addition, we provide a comprehensive analysis of the substitutions on a semantic level which includes the effect of  phonetic\footnote{https://en.wikipedia.org/wiki/Help:IPA/Arabic}, morphological and syntactic  changes illustrated in Table \ref{ling_analysis}. 
\begin{table}[!ht]
    \caption{Linguistic analysis of the most common substitutions from Table \ref{substitusions} for both E2E Transformer and the linguist. The assessment includes phonetic, morphological and syntactic changes}
    \label{ling_analysis}
    \resizebox{\textwidth}{!}{%
    \begin{tabular}{|c|c|c|c|l|}
      \toprule 
      \textbf{Arabic } &  \textbf{BW} & \textbf{Base Phrases} & Phonetic & Meaning / Analysis
      \\
      \hline
      \< إنه / أنه>& $<$nh /  & $<$n+\textcolor{red}{h} / & \textipa{P}\textcolor{red}{i}nnahu /  & h = affix, refers to "him", both prepositions can be used \\ 
      & $>$nh &$>$n+\textcolor{red}{h} &\textipa{P}\textcolor{red}{a}nnahu & for emphasis, their usage and meaning differ based on the syntax 
      \\ 
      \hline
      \< الدعوى / الدعوة> & AldEwY /  & \textcolor{red}{Al}+dEwY / & alda\textipa{Q}wa: / & Al = the; dEwY (root: Ada{\textasciitilde}EY)= To claim something (in court) \\
& AldEwp & \textcolor{red}{Al}+dEwp & alda\textipa{Q}wa: & dEwp (root: dEA)= asks or calls (A man asks/calls God for help)  
 \\ 
 \hline
      \<الجابري / الجبر> & AljAbry /  & \textcolor{red}{Al}+jAbry /  & al{\textdyoghlig}a:b\textcolor{red}{iri}: /  &  Al= the \\
&Aljbr& \textcolor{red}{Al}+jbr & al{\textdyoghlig}abr & AljAbry (root: jbr) = Algebra (Name of a person)
      \\ 
      \hline
     \< دياب / الدياب>  & dyAb / & dyAb / &di:a:b /& Al= the \\
     & AldyAb  & \textcolor{red}{Al}+dyAb &   \textcolor{red}{al}di:a:b & dyAb (dialect of *\}Ab)= wolves (Name of a person) 
     \\ \hline
     \< اللي / التي>   & Ally / & Ally / & a:lla\textcolor{red}{jj} /& Ally = which (dialectal form)
  \\
     & Alty  & Alty &   a:llat\textcolor{red}{i:} & Alty = Alty: which
     \\ 
     \hline
     \< لأن / لئن>   & l$>$n / & \textcolor{red}{l}+$>$n / & l\textipa{P}\textcolor{red}{a}n /  & l =  affix used for explanation; l$>$n (root: $>$n) = to
  \\
     &  l\}n  & \textcolor{red}{l}+\}n &  l\textipa{P}\textcolor{red}{i}n & l\}n (root: \}n) = if
     \\ 
     \hline
     \< للازالت / زالت>   & lAzAlt / &\textcolor{red}{l}+zAlt  / & \textcolor{red}{la}:za:lat /  & lA = affix used for negation
  \\
     & zAlt  & zAlt & za:lat & zAlt (root: zAl) = gone away
     \\
      \bottomrule
    \end{tabular}}
\end{table}
We note that one representative example was picked from Table \ref{substitusions} for each case described in Table \ref{ling_analysis}.\\
\textbf{Phonetic:} It can be seen that phonetic changes are mainly related to the dialectal substitutions (nHnA / nHn, Ally / Alty, AlmAyh / AlmwyA), which does not effect the meaning of the words, or with unfamiliar names (AljAbry / Aljbr). In addition, it is noticeable that, unlike the linguist, the model fails to correctly transcribe words that sound the same but written differently and have different meanings (AldEwY / AldEwp).\\
\textbf{Morphological:} morphological substitutions in some cases cause a change in the word meaning like discarding the negation in (lAzAlt / zAlt). However other changes like adding or discarding a definite article like (al = the), which is commonly used in Arabic language with entities (AlwA\$nTn / wA\$nTn, dyAb / AldyAb), does not affect the meaning.\\
\textbf{Syntactic:} It can be noted that both the model and the linguist find it difficult to correctly transcribe ($<$nh / $>$nh) which are often wrongly used interchangeably by native Arabic speakers. Both ($<$nh / $>$nh) can be used for emphasis, however their usage and meaning differ based on the syntax. In addition, native speakers articulate the first sound (\textipa{P}i ,\textipa{P}a) with a sound that is in between which makes it harder to distinguish. On the other hand, the substitution (l$>$n / l\}n) changes the meaning and one can note that this type of mistake is only made by the machine. Finally, from Table \ref{insertion_deletion} it can be seen that the insertion and deletion patterns are similar for both the linguist and E2E: prepositions are the most frequent errors. It can be observed that the deletion rate of E2E compared to the linguist is somewhat higher than the substitution and insertion rates. This makes sense as the linguist is more careful in transcribing everything that is being heard. 
\begin{table}[!ht]
  
    \caption{Most common insertion and deletions for E2E ASR system and expert linguist.}
    \label{insertion_deletion}
    \centering
    \resizebox{\textwidth}{!}{%
    \begin{tabular}{|c|c|c|c||c|c|c|c|}
    
      \toprule 
      \multicolumn{4}{|c||}{\textbf{Insertions}} &  \multicolumn{4}{c|}{\textbf{Deletions}}\\
     \hline 
      \multicolumn{2}{|c|}{\textbf{E2E Transformer}} &  \multicolumn{2}{c||}{\textbf{linguist}} & \multicolumn{2}{c|}{\textbf{E2E Transformer}} &  \multicolumn{2}{c|}{\textbf{linguist}}\\ 
     \hline
     
      BW & Translation &  BW & Translation &  BW & Translation &  BW & Translation\\ 
      \hline
      \colorbox{pink}{20: $>$n} & that & \colorbox{pink}{16: fy}  & in & \colorbox{yellow}{42: yEny} & means & \colorbox{yellow}{21: nEm} & yes \\
      \colorbox{pink}{18: $>$w} & or & \colorbox{pink}{12: mn} & from & \colorbox{yellow}{39: nEm} & yes & \colorbox{yellow}{17: fy} & in \\ 
        13: Alh & his  & 12: nEm  & yes & \colorbox{yellow}{16: $>$n} & that & \colorbox{yellow}{13: $>$n} & that \\ 
        \colorbox{pink}{12: fy} & in & \colorbox{pink}{10: $>$n} & that & \colorbox{yellow}{16: fy} & in & 10: mA & what\\
     \colorbox{pink}{9: lA} & no & \colorbox{pink}{9: mA} & what & 16: bn & son & \colorbox{yellow}{9: lA} & no\\
     \colorbox{pink}{ 8: mA} & what & 9: w & and  & \colorbox{yellow}{13: mn} & from & 6: Al|n & now \\
      \colorbox{pink}{6: $<$lY} & to & \colorbox{pink}{6: $<$lY} & to  & 13: w & and & \colorbox{yellow}{6: mn} & from\\
      \colorbox{pink}{5: mn} & from & \colorbox{pink}{6: $>$w} & or & \colorbox{yellow}{12: lA} & no & \colorbox{yellow}{6: hw} & him\\
      5: hw & him & 6: yEny &  means & 10: Tyb & alright & \colorbox{yellow}{6: yEny} &  means\\
      4: $>$nA & me & \colorbox{pink}{5: lA}  & no & \colorbox{yellow}{11: hw} & him & 5: yA & (calling)\\
       
      \bottomrule
    \end{tabular}}
\end{table}

\subsection{Speech recognition by E2E transformer and HMM-DNN}

In this section, the results of both the HMM-DNN and the E2E transformer are analyzed, pointing out the  advantages and disadvantages of each. The correct transcription in the examples are highlighted in green and the corresponding errors are highlighted in pink.
\begin{itemize}
    \item \textbf{Dialectal Arabic}: in the case of dialects and overlapped speech, the E2E generates more accurate transcriptions. The E2E transformer is able to learn the context much better with the self-attention mechanism and capture the semantics in both the standard Arabic structure as well as different Arabic dialects.  
    
    \underline{\textit{REF}:} \< التقسيط في شيء سلبي وفي شيء إيجابي هلأ في شيء ضروريات مثلا للبيت>
    
    \underline{\textit{REF\_BW}:}
    AltqsyT \colorbox{green}{fy \$y' 
    slby} wfy \colorbox{green}{\$y' 
    $<$yjAby} hl$>$ \colorbox{green}{fy 
    \$y' DrwryAt} \\
    \colorbox{green}{mvlA llbyt}.
    
    \underline{\textit{Translation}:} The installment has negative things and positive things now there are necessities, for example for a house.
    
    \underline{\textit{E2E}:} \colorbox{green}{fy \$y' slby} \colorbox{pink}{fy}  \colorbox{green}{\$y' $<$yjAby} \colorbox{pink}{$>$nA} \colorbox{green}{fy \$y' DrwryAt mvlA llbyt}
    
    \underline{\textit{HMM-DNN}:} \colorbox{pink}{hy t$>$Syl b\$y'} \colorbox{green}{slby} \colorbox{pink}{b\$y' $<$yjAby Drwryp tsll Albyt} 
    \\
    
    
    
   

    
    \item \textbf{Noise and hesitation}: the E2E transformer is more sensitive to noise and hesitation compared to HMM-DNN. For example, when a person says "aaaah", or when there is a sound of an ambulance, the E2E model generates random words. We think that HMM-TDNN is more robust to noise and hesitation because it uses the $<$SIL$>$ symbol  in the Viterbi alignment during the training step which was successfully aligned with silences. On the other hand, the CTC loss in E2E uses a similar symbol  for the white spaces between the words, however we think that the E2E requires more data or explicit labeling for the noise and hesitation to improve the robustness. This is mainly because the E2E system optimization is mainly driven by the data compared to the HMM-DNN system which includes more expert engineering.
    
    \underline{\textit{REF}:} $<$ambulance\_noise$>$ \<   الفيديو بعد ما انتشر بهيك سرعة يعني أعطاني دافع إني>\\
\< أصير هيك كل أحداث>   \\ 
    \underline{\textit{REF\_BW}:} \colorbox{cyan}{$<$ambulance\_noise$>$} \colorbox{green}{Alfydyw} bEd mA Ant\$r bhyk srEp   yEny $>$ETAny dAfE $<$ny $>$Syr hyk kl $>$HdAv  
    
    \underline{\textit{Translation}:} The video after it spread with such speed. I mean it gave me an incentive that I would like that all events.
   
    
    \underline{\textit{E2E}:}  \colorbox{pink}{fy \$hr mAyw fy \$hr mAyw fy}  bEd mA Ant\$r bhyk bsrEp yEny $>$ETAny dAfE Eny $>$Syr hyk kl $>$HdAv
    
    
    \underline{\textit{HMM-DNN}:} \colorbox{green}{Alfydyw} bEd mA Ant\$r bhyk srEp yEny $>$ETAny dAfE $<$ny $>$Syr hyk kl $>$HdAv

\end{itemize}


\subsection{Effect of data size on ASR performance}

\begin{figure}[!ht]
\begin{center}
\includegraphics[width=7cm,height=4cm]{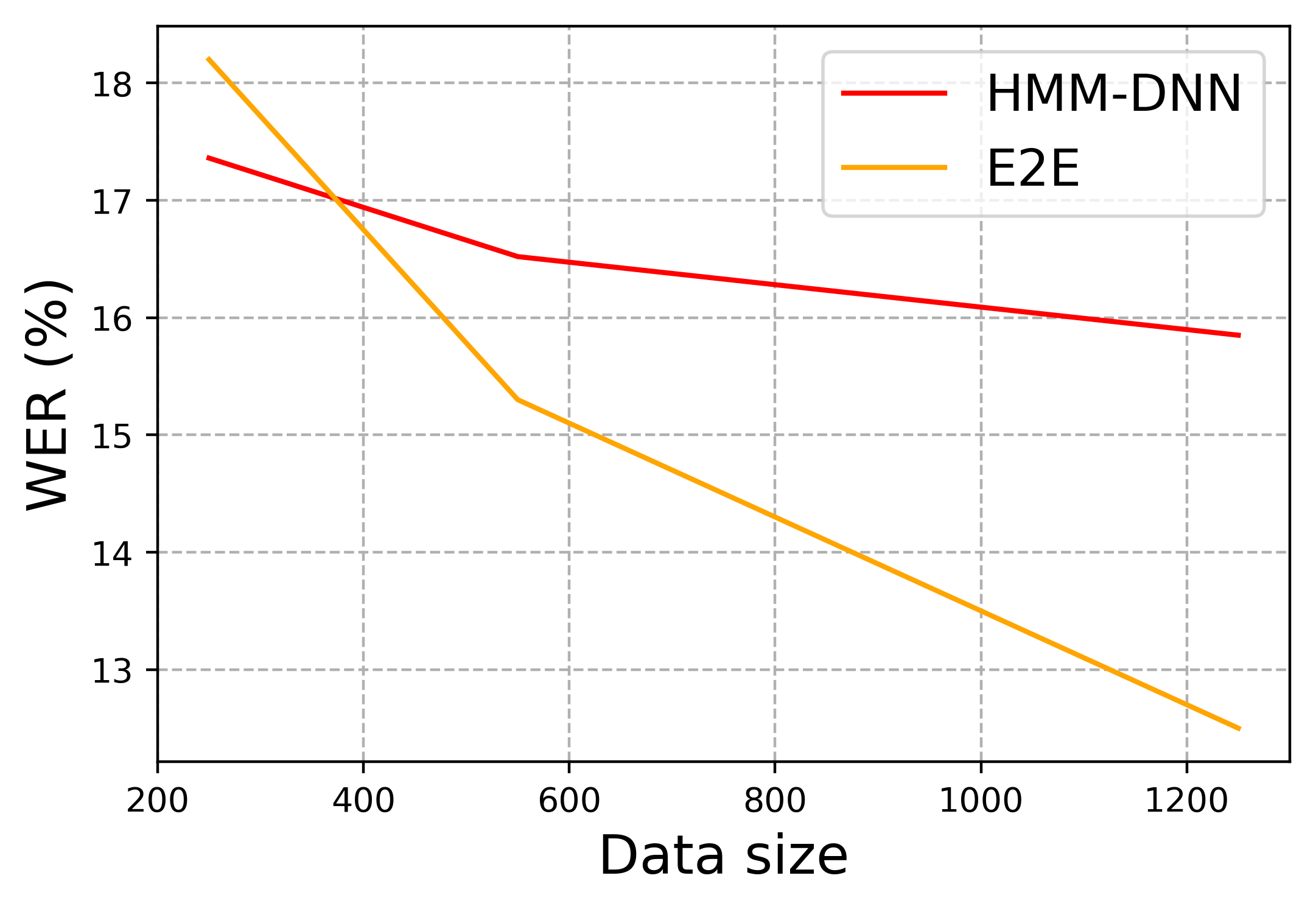}
\caption{Performance of HMM-DNN and E2E ASR systems measured in WER for different training datasizes.}
\label{Datasize_fig} 
\end{center}
\end{figure}

In this section, the effect of the size of the data on both our proposed E2E Transformer and HMM-DNN modular systems is examined. The configuration used for both E2E Transformer and HMM-DNN modular systems is described in Section \ref{hyperparameter_tuning}. The size of the training data was chosen from the following points (250h, 550h, and 1200h). For consistency, the development and testing data were kept the same for all training data sizes. The performance of both E2E and modular systems for each data size is illustrated in Figure \ref{Datasize_fig}.
It can be seen that the performance of the modular ASR system is much better at datasize lower than around 400h. However, as the datasize increases, the E2E performance improves with much steeper trend compared to the modular system, which indicates that with more data, the E2E model is expected to show further improvements. In addition, the results show that the E2E  performance outperforms the modular system after around 400h of data size, which dispel the myth about the need of a very huge amount of private data to match the performance of a modular system \cite{zeyer2018improved}. We note that while 400 hours might not seem a lot of data for English, it is still a huge amount of transcribed data for the vast majority of languages and dialects in the world.

\subsection{E2E Transformer LM ablation analysis}\label{E2E ablation analysis}
In this section, the effect of the LM on the acoustic E2E transformer performance using the transcription text (TR) and background (BG) text is investigated. The WER\% and the real time factor (RT) of the acoustic E2E transformer with RNN LM and Transformer LM architectures are summarized in Table \ref{E2E_results}. 
\begin{table}[!ht]
    \caption{Comparison (WER\% and RT factor) of E2E Transformer (E2E-T) with different LM rescoring configuration on MGB2 test set. The language models were trained with the transcription text (TR) and background (BG) text}
    \label{E2E_results}
    \centering
    \resizebox{0.8\textwidth}{!}{
    \begin{tabular}{lrrrrrr}
      \toprule 
      \multirow{ 2}{*}{\textbf{Method}} & \multicolumn{2}{l}{\textbf{Beam 20}} & \multicolumn{2}{l}{\textbf{Beam 5}}&  \multicolumn{2}{l}{\textbf{Beam 2}} \\ \cline{2-7}
      & WER & RT  & WER & RT  & WER & RT   \\ 
      
      
      E2E-T$+$LSTM-LM (BG+TR)  & 13.4 & 3.65 & 14.6  & 0.9 & 14.7 & 0.44\\

      
      E2E-T$+$T-LM (BG+TR) & 12.7 & 5.87 & 13.1 & 1.49 & 13.5& 0.62 \\ 
      
       E2E-T$+$T-LM (TR) & 12.6 & 5.87 & 13 & 1.49 & 13.4 & 0.62\\
       
      
      E2E-T & \textbf{12.5} & \textbf{3.24} & \textbf{13} & \textbf{0.82} & \textbf{13} & \textbf{0.39}  \\
       
      \bottomrule
    \end{tabular}}
\end{table}
\begin{table}[!ht]
  \centering
    \caption{LM standardized confusion pairs}
    \label{E2E_confusion}
    \resizebox{.6\textwidth}{!}{%
    \begin{tabular}{c|c|c|c|c|c}
      \toprule 
      \textbf{Freq} & \multicolumn{2}{c|}{\textbf{E2E-T}} &  \multicolumn{2}{c|}{\textbf{E2E-T$+$ LM}} & \textbf{Translation} 
      \\ 
       \hline
     {19}& {lly} & {\<اللي> } & {Alty/Al*y} & {\<الذي>/ \<التي>} & {which}
      \\ 
       \hline
    18 &  dA/dh & \<ده> / \<دا> & h*A & \< هذا> & this
    \\
     \hline
     
     10 &  dy/hAy & \<دي> / \<هاي>& h*h & \< هذه> & this 
     \\
     \hline
     
 
    3 & $<$HnA & \<إحنا>  & nHn &\<نحن> & we 
     \\
      \bottomrule
    \end{tabular}}
\end{table}  
It can be seen that as the beam size decreases, the WER gets worse and the RT improves, which is expected. A very interesting aspect to note is that the E2E model without LM achieves better WER compared to E2E with LM. Looking at the confusion results, we found that the LM  rescoring is in fact standardizing the Arabic dialects text to MSA as shown in Table \ref{E2E_confusion}. Although these predictions are linguistically correct, they are counted as errors since they do not match the verbatim transcription, causing the WER with LM to become slightly higher. Overall, the manual inspection showed that the results with LM helps to improve the quality of the transcription without changing its meaning.
\begin{figure}[!ht]
\begin{center}
\includegraphics[width=10cm,height=4.5cm]{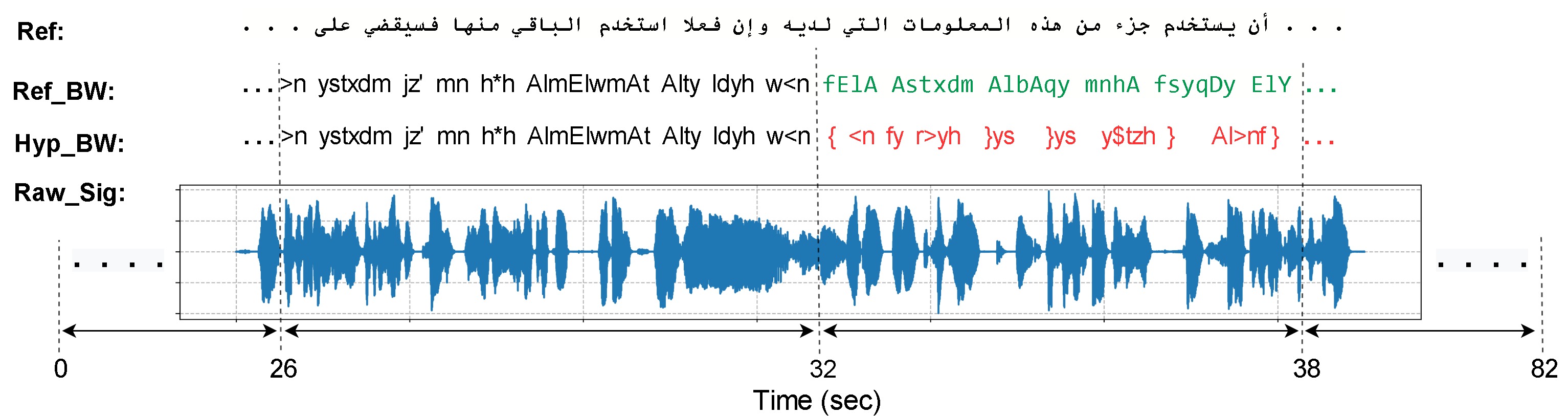}
\caption{Illustration of E2E transformer ASR prediction on very long segment}
\label{long_seg} 
\end{center}
\end{figure}
\subsection{Effect of speech segments length}\label{VAD}
\begin{figure*}[!ht]
     \subfloat[\label{dist0}]{%
     \centering
      \includegraphics[width=5cm,height=3.8cm]{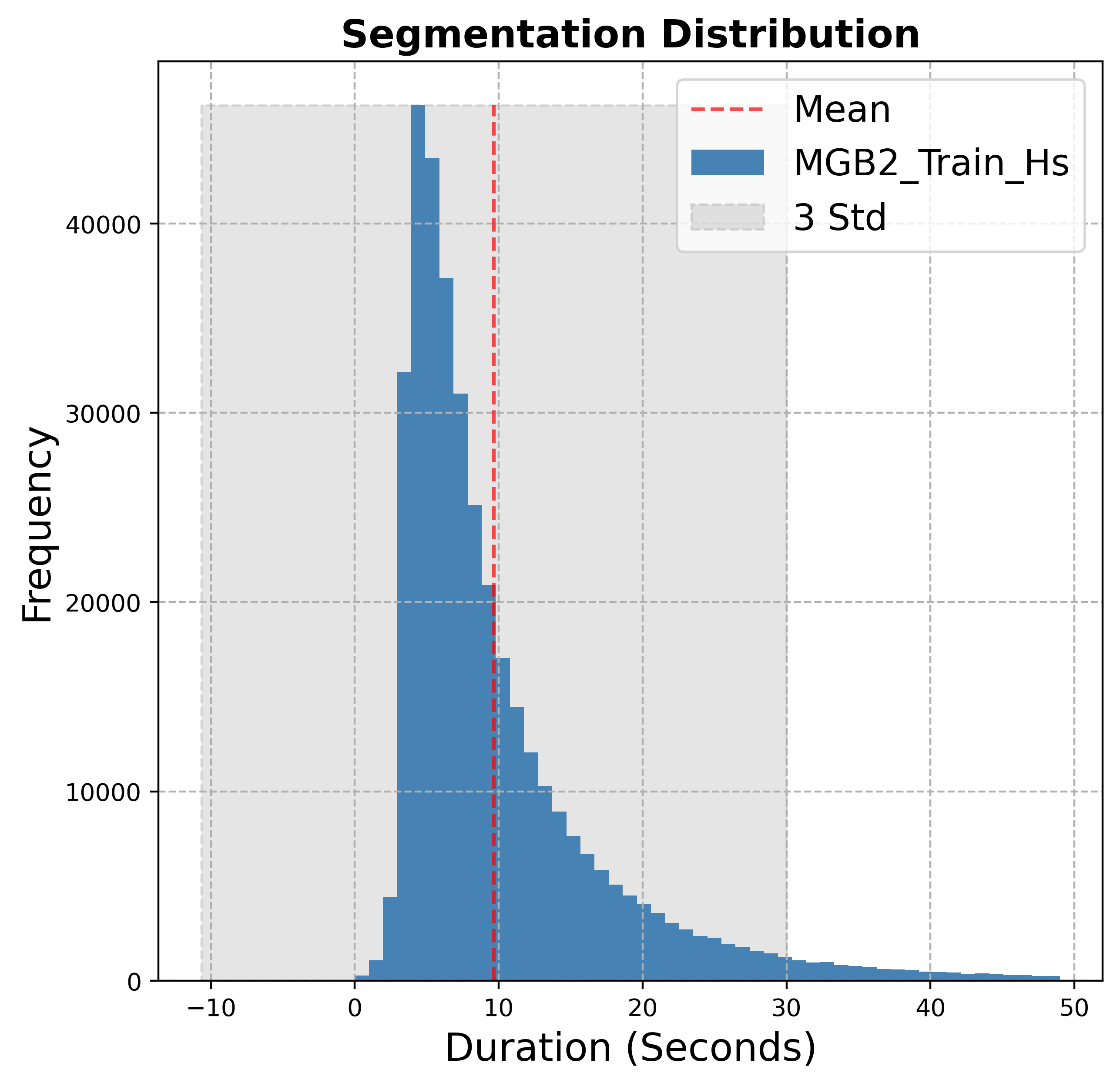} }
      \hspace{\fill}
    \subfloat[\label{dist1}]{%
    \centering
      \includegraphics[width=5cm,height=3.8cm]{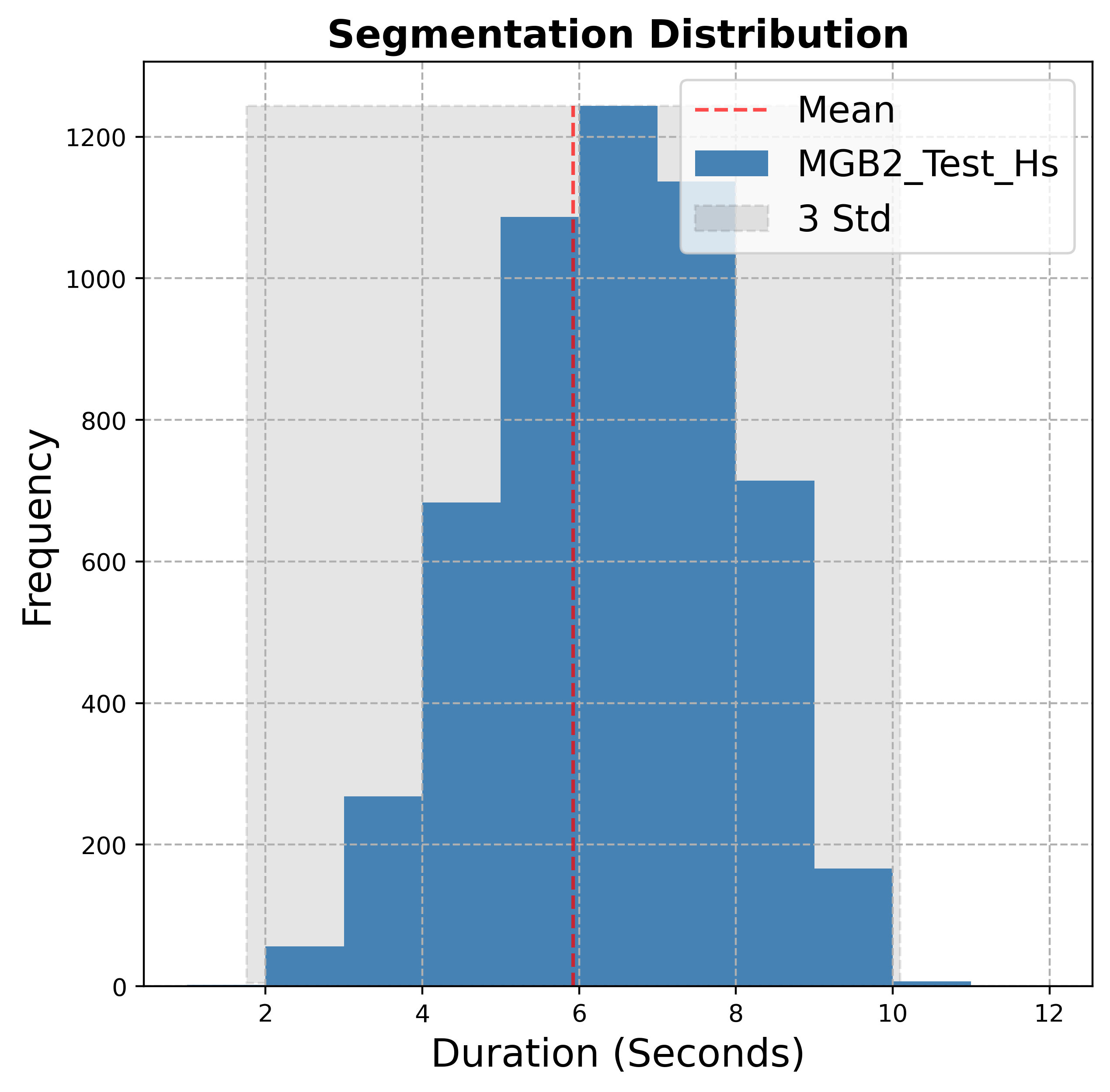} }
      \hspace{\fill}
    \subfloat[\label{dist2}]{%
    \centering
      \includegraphics[width=5cm,height=3.8cm]{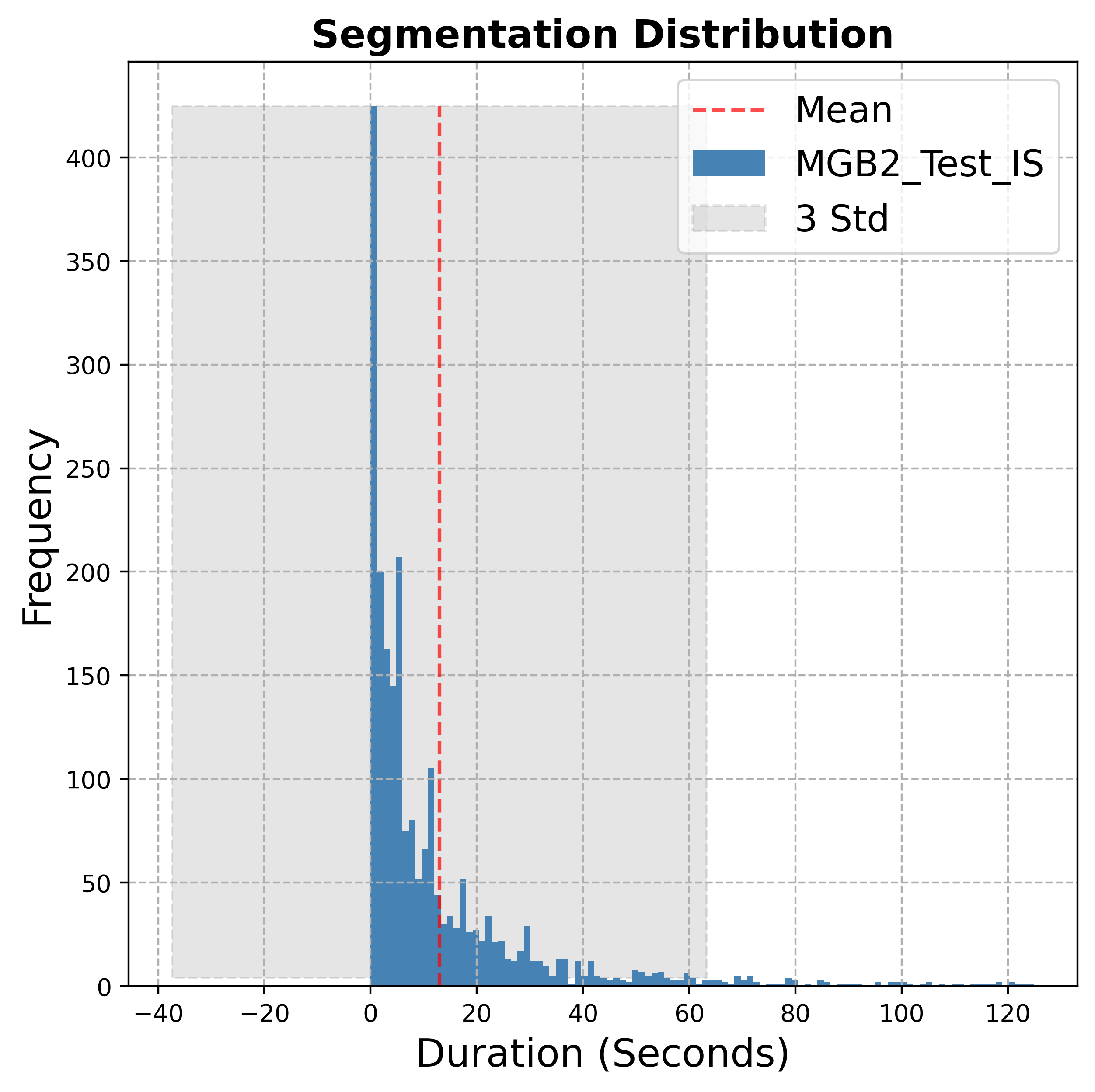} }
    \hspace{\fill}
    \subfloat[\label{dist3}]{%
    \centering
      \includegraphics[width=5cm,height=3.8cm]{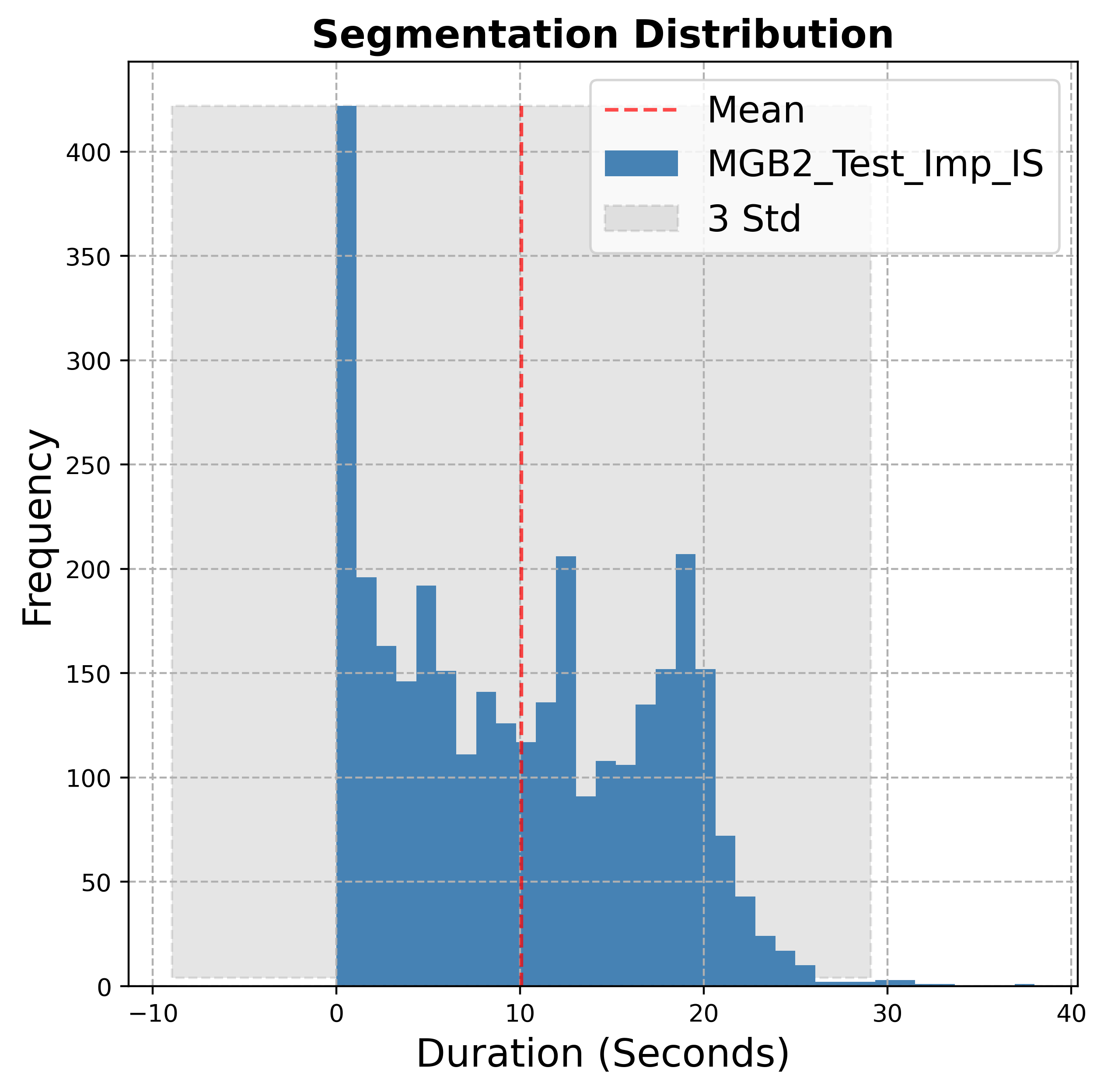} }
    \caption{\small{ \textit{Duration distributions of the speech segments for: (a) MGB2\_Train\_Hs human segmentation, (b) MGB2\_Test\_Hs human segmentation, (c) MGB2\_Test\_IS InaSeg segmentation and (d) MGB2\_Test\_Imp\_IS proposed improved InaSeg segmentation.} 
    }}%
    \label{duration_dist}%
\end{figure*}

The duration of speech segments for ASR has traditionally been controlled by professional transcribers. In practical ASR applications, the duration variability arises from automatic segmentation. In this section, we investigate the impact of segment duration variability on the E2E transformer (E2E-T) and HMM-DNN system that arises from using voice activity detection (VAD). In a practical setup, we found that automatic VAD like InaSpeechSegmenter (IS) \cite{ddoukhanmirex2018} can produce very long segments compared to the training data prepared by the expert human. The change in the segment duration affects the statistical properties of the data and hence causes a shift in the data distribution. An exmaple that shows the effect of a very long segment on E2E transformer transcription is illustrated in Figure \ref{long_seg}. In this particular example, the WER was \%3 up to 32 seconds. However, after 32 seconds, the model starts to generate random predictions highlighted in red. 
To address this problem, we propose improved VAD pipeline (Imp\_IS) that combines benefits of IS, which is accurate in speech detection, and energy-based VAD\footnote{https://github.com/shammur/pyVAD} to control the maximum segment duration. The Imp\_IS is applied by running energy-based VAD and IS in parallel, hence the approach does not introduce any further latency. After that, the resulted segments from energy-based VAD are aligned with the corresponding long segments from IS. In this work, we found that the best results of the proposed Imp\_IS VAD were obtained with the maximum duration threshold of 25 sec. It is worth noting that the automatic segmentation is applied on the complete audio files, and hence the entire MGB2 test set with overlapped segments is used in Tables \ref{seg_wer1},\ref{seg_wer2} compare to the official MGB2 test set used previously in Tables \ref{soa_results}, \ref{E2E_results} that only includes the non-overlapped segments. The WER(\%) of E2E-T and HMM-DNN results on human segmentation HS, IS segmentation, and the proposed Imp\_IS are summarized in Table \ref{seg_wer2}.
\begin{table}[!ht]
\begin{center}
\caption{E2E transformer (E2E-T) and HMM-DNN WER (\%) results on human segmentation (HS), InaSpeech (IS) segmentation , and the proposed improved InaSpeech segmentation (Imp\_IS) benchmarked on two sets: MGB2\_Test, and new Hidden test set (Hidden\_Test) described in Section \ref{Hidden_Test}.}
\label{seg_wer2}
\resizebox{.6\textwidth}{!}{%
\begin{tabular}{|c|c c c|c c c|}
\hline

 & \multicolumn{3}{c|}{\textbf{MGB2\_Test}} & \multicolumn{3}{c|}{\textbf{Hidden\_Test}}\\
  \cline{2-7}
  & HS  & IS & Imp\_IS & HS  & IS & Imp\_IS \\ \toprule
E2E-T  & \textbf{14.7} & 42.3 & \textbf{16.6} & \textbf{12.6} & 32.2 & \textbf{14.5} \\
  \hline
  HMM-DNN & 21.6 & \textbf{27.3} &  28.3 & 15.9  & \textbf{24.8} & 26.6 \\
  
  \bottomrule
  
\end{tabular}}
\end{center}
\end{table}
 It is noticeable that the proposed Imp\_IS approach improved IS WER by an absolute of 25.7\% and 17.7\% on MGB2\_Test and Hidden\_Test respectively, reaching ideal human segmentation with an absolute difference of only 1.9\%. On the other hand, the effect of the segment duration on HMM-DNN is 1.4\% on average when comparing the IS to Imp\_IS segmentation WER results. However, the HMM-DNN results are still affected by the automatic segmentation with absolute difference of 6.3\% and 8.9\%  when comparing HS with Imp\_IS. Moreover to examine the effect of the segmentation duration the WER results of E2E-T and HMM-DNN on the MGB2-Test set with segmentation ranges of 0-15 seconds, 16-30 seconds and 30-more seconds are illustrated in Table \ref{seg_wer1}.
\begin{table}[!ht]
\begin{center}
\caption{E2E transformer (E2E-T) and HMM-DNN WER (\%) results on the MGB2-Test set with segmentation ranges of 0-15 seconds, 16-30 seconds and 30-more seconds.}
\label{seg_wer1}
\resizebox{.5\textwidth}{!}{%
\begin{tabular}{|c | c c c|}
\hline

 & \multicolumn{3}{c|}{\textbf{MGB2\_Test segmentation}}\\
  \cline{2-4}
  & \textbf{0-15 sec}  & \textbf{16-30 sec} & \textbf{30-more sec} \\ \toprule
E2E-T  & \textbf{19.4} & \textbf{18.3} & 59.7  \\
  \hline
  HMM-DNN & 31.7 & 28.3 & \textbf{29.5}  \\
  
  \bottomrule
\end{tabular}}
\end{center}
\end{table}
It can be observed from Table \ref{seg_wer1} that for 30s-more segmentation the WER of E2E-T increases significantly by an absolute of 41.4\% compared to 16-30 sec segmentation range. On the other hand, the WER of HMM-DNN for 30-more segmentation decreased by only 1\% compared to 16-30 sec range. This clearly shows the significant impact of very long segments on the E2E-T system compared to the modular ASR system.
\begin{figure}[!ht]
\begin{center}
\includegraphics[width=10cm,height=5cm]{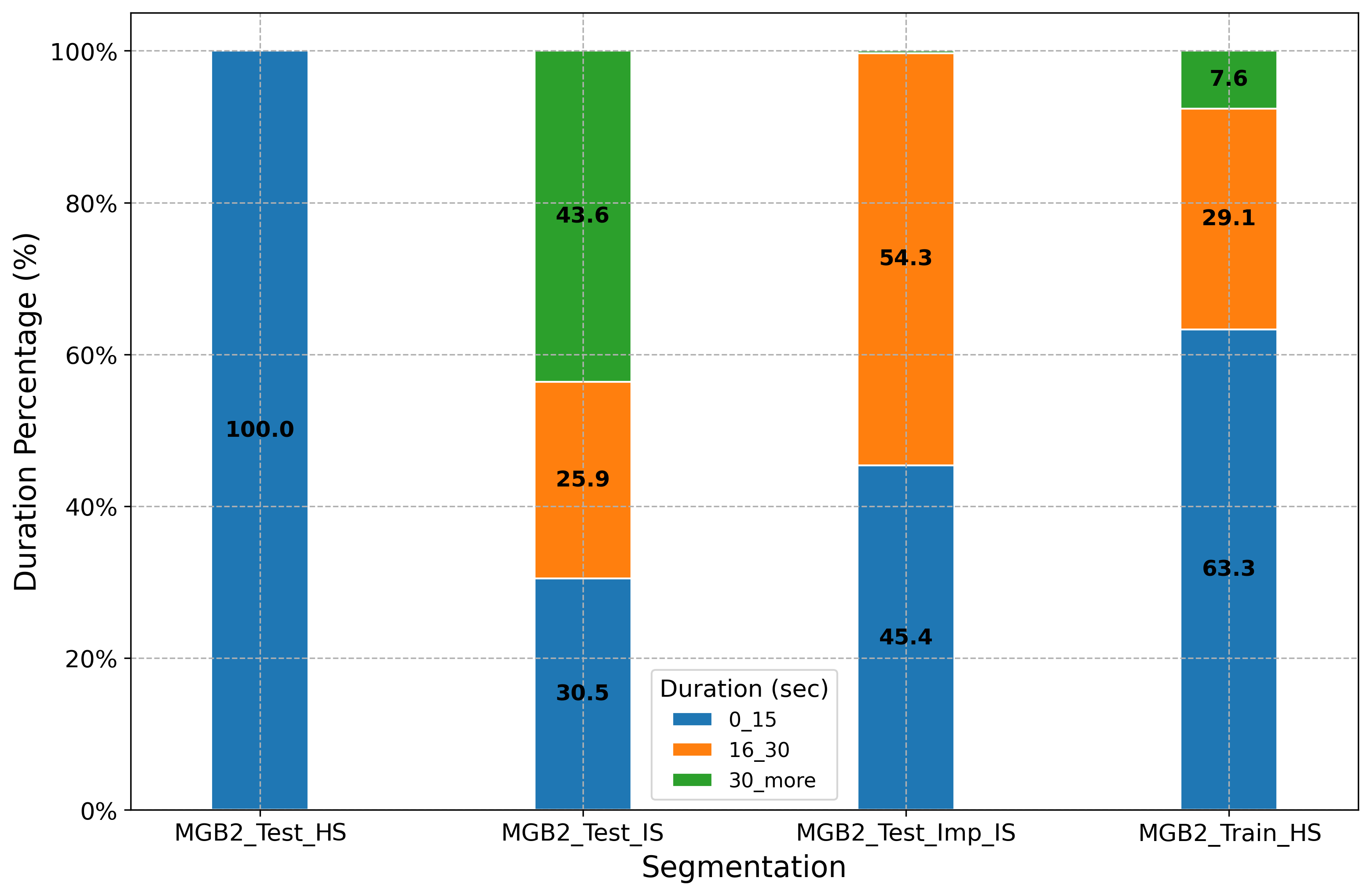}
\caption{Segments duration percentage for different segmentation including MGB2 Train (MGB2\_Train) and Test (MGB2\_Test) human segmentation, InaSeg segmentation (MGB2\_Test\_IS), and improved InaSeg segmentation (MGB2\_Test\_Imp\_IS).}
\label{Segmentation} 
\end{center}
\end{figure}
To further investigate the segmentation duration of the proposed Imp\_IS segmented MGB2\_Test (MGB2\_Test\_Imp\_IS) compared to the human segmented train (MGB2\_Train), human segmented test (MGB2\_Test), IS segmented MGB2\_Test (MGB2\_Test\_IS), we visualize the percentage of segmentation duration with three ranges: 0-15 seconds, 16-30 seconds and 30-more seconds as shown in Figure \ref{Segmentation}.
It can be seen that IS contains 43.6 \% of its segments with duration larger than 30 seconds, which is significantly higher compared to the MGB2\_Train segmentation with only 7.9\% of segments larger than 30 seconds. On the other hand, all segments of the proposed Imp\_IS are less than 30 seconds with 45.4\% less than 15 seconds and 54.3\% between 16 and 30 seconds. Finally, the MGB2 test set with human segmentation has all of its segment duration below 15 seconds. Moreover, we visualize the  segmentation duration statistically we define the effective segmentation duration as the range within 3 standard deviations that covers 99\% of the segments as shown in Figure \ref{duration_dist}. 
It can be seen that the effective segmentation duration of MGB2\_Test\_IS is around 60 seconds with maximum duration of around 120 seconds which is much larger than the effective segmentation duration of MGB2\_Train, which is only 30 seconds. This confirms our assumption about the shift in the duration distribution and explains why the model fails to make successful predictions after around 30 seconds.

\subsection{Dialectal Arabic ASR}
In this section, the performance of E2E transformer is studied on two dialectal Arabic ASR challenges MGB3 and MGB5 \cite{ali2017speech,ali2019mgb}. For more details about the data, refer to section \ref{Hidden_Test} and Table \ref{mgb2_1}. As DA is lacking standard orthographic rules as well as sizable transcribed data, it is considered to be an excellent choice to highlight the challenges of speech recognition in the wild. The best E2E transformer obtained in Section \ref{E2E ablation analysis} is benchmarked with the MGB3  and the MGB5 state-of-the-art results. The E2E transformer was finetuned on the MGB3 and the MGB5 adaptation sets independently. The hyper-parameters for the transformer model are similar to what was used for the MGB2 training described in Section \ref{hyperparameter_tuning} except that the learning rate is reduced to 0.1 and no warm-up steps. The model is initialized with the E2E transformer parameters pretrained on the MGB2 from Section \ref{E2E ablation analysis}. 
For data preprocessing, the same data augmentation described in Section \ref{Data_prep} was followed. Each sentence in the adaptation and the development data were transcribed by four different annotators to explore the non-orthographic nature of dialectal Arabic. In our experiments, the transcripts from the four transcribers were combined, which increased the amount of the data four times.
The participants in the MGB3 and the MGB5 challenges had the access to the development set which often used in the competitions as part of the  model finetuning. As a result we report two sets of results:
1) Finetune the model using the adaptation set only for $30$ epochs and use the development set to monitor the performance and select the model with the best result; 
2) Finetune the model on adaptation  and the development sets combined. In this case the model is first finetuned  for $30$ epochs on the adaptation set and then further finetuned on adaptation set for an additional $5$ epochs. The results of both approaches on the MGB3\_Test, MGB5\_Test sets are illustrated in Table \ref{E2E_mgb3_mgb5}. 
\begin{table}[!ht]
  \begin{center}
    \caption{MR-WER\% \& AV-WER\% results on two dialectal datasets MGB3 and MGB5 test sets.}
    \label{E2E_mgb3_mgb5}
    \resizebox{0.8\textwidth}{!}{%
    \begin{tabular}{|l|c|c|c|c|}
      \toprule 
      \multirow{2}{*}{} & \multicolumn{2}{c|}{\textbf{MGB3\_Test}} & \multicolumn{2}{c|}{\textbf{MGB5\_Test}} \\
      \cline{2-5}
       & MR-WER & AV-WER & MR-WER & AV-WER\\
      
      \hline 
      
      Aalto \cite{smit2017aalto} & 29.3 & 37.5 & - &-\\

      RDI-CU \cite{ali2019mgb} & - & - & 37.6 & 59.4\\

       E2E-Transformer(adapt)  & 29.2 & 36.0 &  34.9
      & 57.2
      \\
      E2E-Transformer(adapt+dev)  & \textbf{27.5} & \textbf{34.4} &  \textbf{33.8}
      & \textbf{56.2}
      \\
      \bottomrule
    \end{tabular}}
  \end{center}
\end{table}
It can be seen from Table \ref{E2E_mgb3_mgb5} that the single E2E-Transformer outperforms the state-of-the-art modular HMM-DNN systems in both of the DA challenges. We see about $4$\% relative reduction in AV-WER while using the adaptation data only, and between $6$-$8$\% relative reduction in AV-WER when using both the adaptation and the development data, which significantly outperforms the previous state-of-the-arts in DA ASR. This is a new milestone for DA speech recognition.

\section{Conclusion}\label{sec_conclusion}
In this paper, we presented the first comprehensive study comparing head to head E2E ASR, modular HMM-DNN ASR and HSR on Arabic speech. We provided a comprehensive error analysis comparing the best ASR system performance to the expert linguist and native speaker. It has been found that the machine ASR arguably outperforms the performance of the native speaker, however, the WER gap to reach expert linguist performance is still on average $3.5$\% on the raw Arabic transcription text. It was noticeable that the machine mistakes showed high similarity with the expert linguist transcription. Additionally, we developed the first E2E transformer for the Arabic ASR and its dialects. The proposed E2E transformer significantly outperformed prior state-of-the-art on MGB2, MGB3 and MGB5 achieving a new
state-of-the-art performance at $12.5$\%, $27.5$\% and $33.8$\% respectively. Moreover, it has been found that, in practical ASR, the segment duration has a severe impact on E2E transformer performance. To address the problem of segment duration variability, a new VAD pipeline with a maximum duration threshold was proposed. For future work, we plan to address the gap between human and machine in Arabic ASR and address the low resource challenge in dialectal Arabic, which still shows a high error rate.

\bibliography{mybibfile}

\begin{thebibliography}{10}
\expandafter\ifx\csname url\endcsname\relax
  \def\url#1{\texttt{#1}}\fi
\expandafter\ifx\csname urlprefix\endcsname\relax\def\urlprefix{URL }\fi
\expandafter\ifx\csname href\endcsname\relax
  \def\href#1#2{#2} \def\path#1{#1}\fi

\bibitem{deepspeech2}
D.~Amodei, S.~Ananthanarayanan, R.~Anubhai, J.~Bai, E.~Battenberg, C.~Case,
  J.~Casper, B.~Catanzaro, Q.~Cheng, G.~Chen, et~al., Deep speech 2:
  {E}nd-to-end speech recognition in {E}nglish and {M}andarin, in:
  International conference on machine learning, 2016, pp. 173--182.

\bibitem{graves2014towards}
A.~Graves, N.~Jaitly, Towards end-to-end speech recognition with recurrent
  neural networks, in: International conference on machine learning, 2014, pp.
  1764--1772.

\bibitem{xiong2016achieving}
W.~Xiong, J.~Droppo, X.~Huang, F.~Seide, M.~Seltzer, A.~Stolcke, D.~Yu,
  G.~Zweig, Achieving human parity in conversational speech recognition, arXiv
  preprint arXiv:1610.05256.

\bibitem{saon2017english}
G.~Saon, G.~Kurata, T.~Sercu, K.~Audhkhasi, S.~Thomas, D.~Dimitriadis, X.~Cui,
  B.~Ramabhadran, M.~Picheny, L.-L. Lim, et~al., English conversational
  telephone speech recognition by humans and machines, Proc. Interspeech (2017)
  132--136.

\bibitem{michalowski2006lives}
P.~Michalowski, The lives of the sumerian language, Margins of writing, origins
  of cultures (2006) 159--84.

\bibitem{smit2017aalto}
P.~Smit, S.~R. Gangireddy, S.~Enarvi, S.~Virpioja, M.~Kurimo, Aalto system for
  the 2017 {A}rabic multi-genre broadcast challenge, in: IEEE Automatic Speech
  Recognition and Understanding Workshop (ASRU), IEEE, 2017, pp. 338--345.

\bibitem{mubarak2019highly}
H.~Mubarak, A.~Abdelali, H.~Sajjad, Y.~Samih, K.~Darwish, Highly effective
  arabic diacritization using sequence to sequence modeling, in: Proceedings of
  the 2019 Conference of the North American Chapter of the Association for
  Computational Linguistics: Human Language Technologies, Volume 1 (Long and
  Short Papers), 2019, pp. 2390--2395.

\bibitem{das2015cross}
A.~Das, M.~Hasegawa-Johnson, Cross-lingual transfer learning during supervised
  training in low resource scenarios, in: Sixteenth Annual Conference of the
  International Speech Communication Association, 2015.

\bibitem{khurana2019darts}
S.~Khurana, A.~Ali, J.~Glass, Darts: Dialectal {A}rabic transcription system,
  arXiv preprint arXiv:1909.12163.

\bibitem{ahmed2018end}
A.~Ahmed, Y.~Hifny, K.~Shaalan, S.~Toral, End-to-end lexicon free {A}rabic
  speech recognition using recurrent neural networks, Computational
  Linguistics, Speech And Image Processing For Arabic Language 4 (2018) 231.

\bibitem{kudo2018subword}
T.~Kudo, Subword regularization: Improving neural network translation models
  with multiple subword candidates, arXiv preprint arXiv:1804.10959.

\bibitem{khurana2016qcri}
S.~Khurana, A.~Ali, Qcri advanced transcription system ({QATS}) for the
  {A}rabic multi-dialect broadcast media recognition: {MGB}-2 challenge, in:
  IEEE Spoken Language Technology Workshop (SLT), IEEE, 2016, pp. 292--298.

\bibitem{ddoukhanmirex2018}
D.~Doukhan, E.~Lechapt, M.~Evrard, J.~Carrive, Ina’s mirex 2018 music and
  speech detection system, in: Music Information Retrieval Evaluation eXchange
  (MIREX 2018), 2018.

\bibitem{giannakopoulos2015pyaudioanalysis}
T.~Giannakopoulos, pyaudioanalysis: An open-source python library for audio
  signal analysis, PloS one 10~(12) (2015) e0144610.

\bibitem{dahl2011context}
G.~E. Dahl, D.~Yu, L.~Deng, A.~Acero, Context-dependent pre-trained deep neural
  networks for large-vocabulary speech recognition, IEEE Transactions on audio,
  speech, and language processing 20~(1) (2011) 30--42.

\bibitem{graves2013speech}
A.~Graves, A.-r. Mohamed, G.~Hinton, Speech recognition with deep recurrent
  neural networks, in: IEEE international conference on acoustics, speech and
  signal processing (ICASSP), IEEE, 2013, pp. 6645--6649.

\bibitem{hifny2015unified}
Y.~Hifny, Unified acoustic modeling using deep conditional random fields,
  Transactions on Machine Learning and Artificial Intelligence 3~(2) (2015)
  65--65.

\bibitem{peddinti2015time}
V.~Peddinti, D.~Povey, S.~Khudanpur, A time delay neural network architecture
  for efficient modeling of long temporal contexts, in: Sixteenth Annual
  Conference of the International Speech Communication Association, 2015.

\bibitem{ali2017speech}
A.~Ali, S.~Vogel, S.~Renals, Speech recognition challenge in the wild: {A}rabic
  {MGB}-3, in: IEEE Automatic Speech Recognition and Understanding Workshop
  (ASRU), IEEE, 2017, pp. 316--322.

\bibitem{wang2019overview}
D.~Wang, X.~Wang, S.~Lv, An overview of end-to-end automatic speech
  recognition, Symmetry 11~(8) (2019) 1018.

\bibitem{chan2016listen}
W.~Chan, N.~Jaitly, Q.~Le, O.~Vinyals, Listen, attend and spell: A neural
  network for large vocabulary conversational speech recognition, in: IEEE
  International Conference on Acoustics, Speech and Signal Processing (ICASSP),
  IEEE, 2016, pp. 4960--4964.

\bibitem{chorowski2015attention}
J.~K. Chorowski, D.~Bahdanau, D.~Serdyuk, K.~Cho, Y.~Bengio, Attention-based
  models for speech recognition, Advances in neural information processing
  systems 28 (2015) 577--585.

\bibitem{watanabe2017hybrid}
S.~Watanabe, T.~Hori, S.~Kim, J.~R. Hershey, T.~Hayashi, Hybrid ctc/attention
  architecture for end-to-end speech recognition, IEEE Journal of Selected
  Topics in Signal Processing 11~(8) (2017) 1240--1253.

\bibitem{belinkov122019analyzing}
Y.~Belinkov12, A.~Ali, J.~Glass, Analyzing phonetic and graphemic
  representations in end-to-end automatic speech recognition, Proc.
  Interspeech.

\bibitem{vaswani2017attention}
A.~Vaswani, N.~Shazeer, N.~Parmar, J.~Uszkoreit, L.~Jones, A.~N. Gomez,
  {\L}.~Kaiser, I.~Polosukhin, Attention is all you need, in: Advances in
  neural information processing systems, 2017, pp. 5998--6008.

\bibitem{karita2019comparative}
S.~Karita, N.~Chen, T.~Hayashi, T.~Hori, H.~Inaguma, Z.~Jiang, M.~Someki,
  N.~E.~Y. Soplin, R.~Yamamoto, X.~Wang, et~al., A comparative study on
  transformer vs rnn in speech applications, in: IEEE Automatic Speech
  Recognition and Understanding Workshop (ASRU), IEEE, 2019, pp. 449--456.

\bibitem{wang2020transformer}
Y.~Wang, A.~Mohamed, D.~Le, C.~Liu, A.~Xiao, J.~Mahadeokar, H.~Huang,
  A.~Tjandra, X.~Zhang, F.~Zhang, et~al., Transformer-based acoustic modeling
  for hybrid speech recognition, in: IEEE International Conference on
  Acoustics, Speech and Signal Processing (ICASSP), IEEE, 2020, pp. 6874--6878.

\bibitem{synnaeve2020end}
G.~Synnaeve, Q.~Xu, J.~Kahn, T.~Likhomanenko, E.~Grave, V.~Pratap, A.~Sriram,
  V.~Liptchinsky, R.~Collobert, End-to-end {ASR}: from supervised to
  semi-supervised learning with modern architectures, arXiv preprint
  arXiv:1911.08460.

\bibitem{stolcke2017comparing}
A.~Stolcke, J.~Droppo, Comparing human and machine errors in conversational
  speech transcription, arXiv preprint arXiv:1708.08615.

\bibitem{ali2016mgb}
A.~Ali, P.~Bell, J.~Glass, Y.~Messaoui, H.~Mubarak, S.~Renals, Y.~Zhang, The
  mgb-2 challenge: Arabic multi-dialect broadcast media recognition, in: IEEE
  Spoken Language Technology Workshop (SLT), IEEE, 2016, pp. 279--284.

\bibitem{povey2016purely}
D.~Povey, V.~Peddinti, D.~Galvez, P.~Ghahremani, V.~Manohar, X.~Na, Y.~Wang,
  S.~Khudanpur, Purely sequence-trained neural networks for {ASR} based on
  lattice-free {MMI}, in: Proc. Interspeech, 2016, pp. 2751--2755.

\bibitem{povey2011kaldi}
D.~Povey, A.~Ghoshal, G.~Boulianne, L.~Burget, O.~Glembek, N.~Goel,
  M.~Hannemann, P.~Motlicek, Y.~Qian, P.~Schwarz, et~al., The kaldi speech
  recognition toolkit, in: IEEE Automatic Speech Recognition and Understanding
  Workshop (ASRU), no. CONF, IEEE Signal Processing Society, 2011.

\bibitem{povey2018time}
D.~Povey, H.~Hadian, P.~Ghahremani, K.~Li, S.~Khudanpur, A time-restricted
  self-attention layer for {ASR}, in: IEEE International Conference on
  Acoustics, Speech and Signal Processing (ICASSP), IEEE, 2018, pp. 5874--5878.

\bibitem{ghahremani2014pitch}
P.~Ghahremani, B.~BabaAli, D.~Povey, K.~Riedhammer, J.~Trmal, S.~Khudanpur, A
  pitch extraction algorithm tuned for automatic speech recognition, in: IEEE
  international conference on acoustics, speech and signal processing (ICASSP),
  IEEE, 2014, pp. 2494--2498.

\bibitem{ba2016layer}
J.~L. Ba, J.~R. Kiros, G.~E. Hinton, Layer normalization, arXiv preprint
  arXiv:1607.06450.

\bibitem{he2016deep}
K.~He, X.~Zhang, S.~Ren, J.~Sun, Deep residual learning for image recognition,
  in: Proceedings of the IEEE conference on computer vision and pattern
  recognition, 2016, pp. 770--778.

\bibitem{ali2019mgb}
A.~Ali, S.~Shon, Y.~Samih, H.~Mubarak, A.~Abdelali, J.~Glass, S.~Renals,
  K.~Choukri, The {MGB}-5 challenge: Recognition and dialect identification of
  dialectal {A}rabic speech, in: IEEE Automatic Speech Recognition and
  Understanding Workshop (ASRU), IEEE, 2019, pp. 1026--1033.

\bibitem{ali2015multi}
A.~Ali, W.~Magdy, P.~Bell, S.~Renais, Multi-reference wer for evaluating asr
  for languages with no orthographic rules, in: IEEE Automatic Speech
  Recognition and Understanding Workshop (ASRU), IEEE, 2015, pp. 576--580.

\bibitem{watanabe2018espnet}
S.~Watanabe, T.~Hori, S.~Karita, T.~Hayashi, J.~Nishitoba, Y.~Unno, N.-E.~Y.
  Soplin, J.~Heymann, M.~Wiesner, N.~Chen, et~al., Espnet: End-to-end speech
  processing toolkit, Proc. Interspeech (2018) 2207--2211.

\bibitem{ali2014complete}
A.~Ali, Y.~Zhang, P.~Cardinal, N.~Dahak, S.~Vogel, J.~Glass, A complete kaldi
  recipe for building arabic speech recognition systems, in: IEEE spoken
  language technology workshop (SLT), IEEE, 2014, pp. 525--529.

\bibitem{ko2015audio}
T.~Ko, V.~Peddinti, D.~Povey, S.~Khudanpur, Audio augmentation for speech
  recognition, in: Sixteenth Annual Conference of the International Speech
  Communication Association, 2015.

\bibitem{park2019specaugment}
D.~S. Park, W.~Chan, Y.~Zhang, C.-C. Chiu, B.~Zoph, E.~D. Cubuk, Q.~V. Le,
  Specaugment: A simple data augmentation method for automatic speech
  recognition, Proc. Interspeech (2019) 2613--2617.

\bibitem{pappagari2019hierarchical}
R.~Pappagari, P.~Zelasko, J.~Villalba, Y.~Carmiel, N.~Dehak, Hierarchical
  transformers for long document classification, in: IEEE Automatic Speech
  Recognition and Understanding Workshop (ASRU), IEEE, 2019, pp. 838--844.

\bibitem{zeyer2018improved}
A.~Zeyer, K.~Irie, R.~Schl{\"u}ter, H.~Ney, Improved training of end-to-end
  attention models for speech recognition, arXiv preprint arXiv:1805.03294.

\end{thebibliography}

\end{document}